\documentclass[pdflatex,sn-mathphys]{sn-jnl}

\usepackage{booktabs}
\usepackage{graphicx}

\theoremstyle{thmstyleone}%
\theoremstyle{thmstyletwo}%

\theoremstyle{thmstylethree}%

\begin{document}

\title[Deep learning for blood vessel segmentation of HiP-CT]{Deep Learning for Vascular Segmentation and Applications in Phase Contrast Tomography Imaging}




\author*[1]{\fnm{Ekin} \sur{Yagis}}\email{e.yagis@ucl.ac.uk}

\author[1,4]{\fnm{Shahab} \sur{Aslani}}
\equalcont{These authors contributed equally to this work.}

\author[3]{\fnm{Yashvardhan} \sur{Jain}}
\equalcont{These authors contributed equally to this work.}

\author[1]{\fnm{Yang} \sur{Zhou}}
\equalcont{These authors contributed equally to this work.}

\author[1]{\fnm{Shahrokh} \sur{Rahmani}}
\equalcont{These authors contributed equally to this work.}

\author[1,2]{\fnm{Joseph} \sur{Brunet}}

\author[5]{\fnm{Alexandre} \sur{Bellier}}

\author[6]{\fnm{Christopher} \sur{Werlein}}

\author[7]{\fnm{Maximilian} \sur{Ackermann}}

\author[8]{\fnm{Danny} \sur{Jonigk}}

\author[2]{\fnm{Paul} \sur{Tafforeau}}
\equalcont{These authors contributed equally to this work.}

\author[1]{\fnm{Peter D} \sur{Lee}}
\equalcont{These authors contributed equally to this work.}

\author[1]{\fnm{Claire} \sur{Walsh}}
\equalcont{These authors contributed equally to this work.}

\affil*[1]{\orgdiv{Department of Mechanical Engineering, University College London, London, UK}}

\affil[2]{\orgdiv{European Synchrotron Radiation Facility, Grenoble, France}}

\affil[3]{\orgdiv{Department of Intelligent Systems Engineering, Luddy School of Informatics, Computing, and Engineering, Indiana University, Bloomington, USA}}

\affil[4]{\orgdiv{Centre for Medical Image Computing, University College London, London UK}}

\affil[5]{\orgdiv{Laboratoire d'Anatomie Des Alpes Françaises, Grenoble, France}}

\affil[6]{\orgdiv{Institute of Pathology, Hannover Medical School, Carl-Neuberg-Straße 1, 30625, Hannover, Germany.}}

\affil[6]{\orgdiv{Institute of Functional and Clinical Anatomy, University Medical Center of the Johannes Gutenberg-University Mainz, Mainz, Germany.}}

\affil[8]{\orgdiv{Member of the German Center for Lung Research (DZL), Biomedical Research in Endstage and Obstructive Lung Disease Hannover (BREATH), Hannover, Germany.}}


\abstract{

Automated blood vessel segmentation is critical for biomedical image analysis, as vessel morphology changes are associated with numerous pathologies. Still, precise segmentation is difficult due to the complexity of vascular structures, anatomical variations across patients, the scarcity of annotated public datasets, and the quality of images.

We present a thorough literature review, highlighting the state of machine learning techniques across diverse organs. Our goal is to provide a foundation on the topic and identify a robust baseline model for application to vascular segmentation in a new imaging modality, Hierarchical Phase-Contrast Tomography (HiP-CT). 
Introduced in 2020 at the European Synchrotron Radiation Facility, HiP-CT enables 3D imaging of complete organs at an unprecedented resolution of ca. 20µm/voxel, with the capability for localized zooms in selected regions down to 1µm/voxel without sectioning. Through our review and application of baseline models to HiP-CT data we aim to provide a future benchmark for models applied to HiP-CT data. We have created a training dataset with double annotator-validated vascular data from three kidneys imaged with HiP-CT in the context of the Human Organ Atlas Project. Finally, utilising the nnU-Net model, we conduct experiments to assess the models' performance on both familiar and unseen samples, employing vessel-specific metrics.

Our results show that while segmentations yielded reasonably high scores—such as clDice values ranging from 0.82 to 0.88, certain errors persisted. Specifically, large vessels that collapsed due to the lack of hydro-static pressure (HiP-CT is an ex vivo technique) were segmented poorly. Moreover, decreased connectivity in finer vessels and higher segmentation errors at vessel boundaries were observed. Such errors, particularly in significant vessels, obstruct the understanding of the structures by interrupting vascular tree connectivity. Through our review and outputs, we aim to set a benchmark for subsequent model evaluations using various modalities, especially with the HiP-CT imaging database.
}




\maketitle

\section{Introduction}\label{sec1}

The vascular, or circulatory system consists of the network of blood vessels responsible for circulating blood, delivering oxygen, and facilitating the removal of waste from tissues and organs \cite{pugsley2000vascular}.
Abnormalities in the structure or function of vascular networks can lead to, or be indicative of, a myriad of pathologies ranging from tumor growth and metastasis to strokes and cardiovascular disorders \cite{lee2019organs}. Hence, accurate segmentation of vasculature and quantitative evaluation of its morphology are widely applied to better understand these pathophysiological processes \cite{potente2017link,sweeney2018role,moccia2018blood,rust2020practical}.

Annotation of images by experts has been regarded as the segmentation gold standard, but it is time-consuming and requires specialised knowledge \cite{zhao2019segmentation}. The success of convolutional neural networks (CNN) in classification leads to researchers attempting to employ deep learning technologies for image segmentation \cite{goni2022brain}. Consequently, a growing number of automated and semi-automatic vessel segmentation methods have been developed over the last years for the diagnosis of diseases associated with the vascular system \cite{fraz2012blood,zhao2019segmentation,moccia2018blood}.

In order to be clinically useful, segmentation algorithms need to be capable of handling a wide range of anatomical/sample variability and to be able to function on different types of imaging modalities and qualities. Advances in 3D imaging, image analysis, and image-based modeling have resulted in complex frameworks for gathering as well as simulating structural and functional physiological data \cite{sweeney2018role}. Magnetic resonance imaging (MRI), positron emission tomography (PET), and computed tomography (CT) have been the three major imaging modalities utilised for decades \cite{goni2022brain}. For vascular imaging, magnetic resonance angiography (MRA) is one of the most common choices due to its noninvasive nature, absence of ionising radiation exposure, capability for non-contrast examination, and capacity to provide volumetric representations that highlight vascular disease \cite{kuo2019vascular}. 

However, the complexity of vascular imaging data and the anatomical variability among subjects make vessel segmentation a challenging task. Despite the clinical need, a fully automated segmentation method for 3D vascular segmentation and subsequent feature extraction has not yet been developed due a number of specific challenges \cite{deshpande2021automatic}. The main hurdles are the variations across scales, differences in individual anatomies, complexities of slender structures, a limited volume ratio (i.e. vessel/background), and a scarcity of publicly available labeled data. In terms of individual anatomical variations, there are significant disparities in vessel length, diameter, and tortuosity, complicating the tasks of vascular tracking and segmentation \cite{deshpande2021automatic,luo2006extraction,lesage2009review,zhao2017automatic}. Additionally, narrower vessels with diameters at the border of the imaging resolution are particularly difficult to capture \cite{deshpande2021automatic,ajam2017review}.


Hierarchical Phase-Contrast Tomography (HiP-CT) \cite{walsh2021imaging} is a nascent imaging technology developed in 2020 on the BM05 bealine at the ESRF. The technique is a propagation-based phase-contrast tomography technique that utilises the increased brilliance and resulting high spatial coherence of the European Synchrotron Radiation Facility’s upgrade to a 4th Generation X-ray source (The Extremely Brilliant Source (EBS). With HiP-CT, it is possible to resolve very small refractive index difference (density difference) in soft tissue structures, even in very large samples such as whole adult human organs. The propagation through free space of the X-rays, after interaction with the sample allow the refractive index shifts to be converted to intensity variation through single-distance phase retrieval at reconstruction \cite{paganin2002simultaneous}. HiP-CT has been used to image whole human organs, down to resolution of some single-cell types in regions of interest \cite{walsh2021imaging}.

\begin{figure*}[!ht]
\centering
  \includegraphics[width=0.8\textwidth]{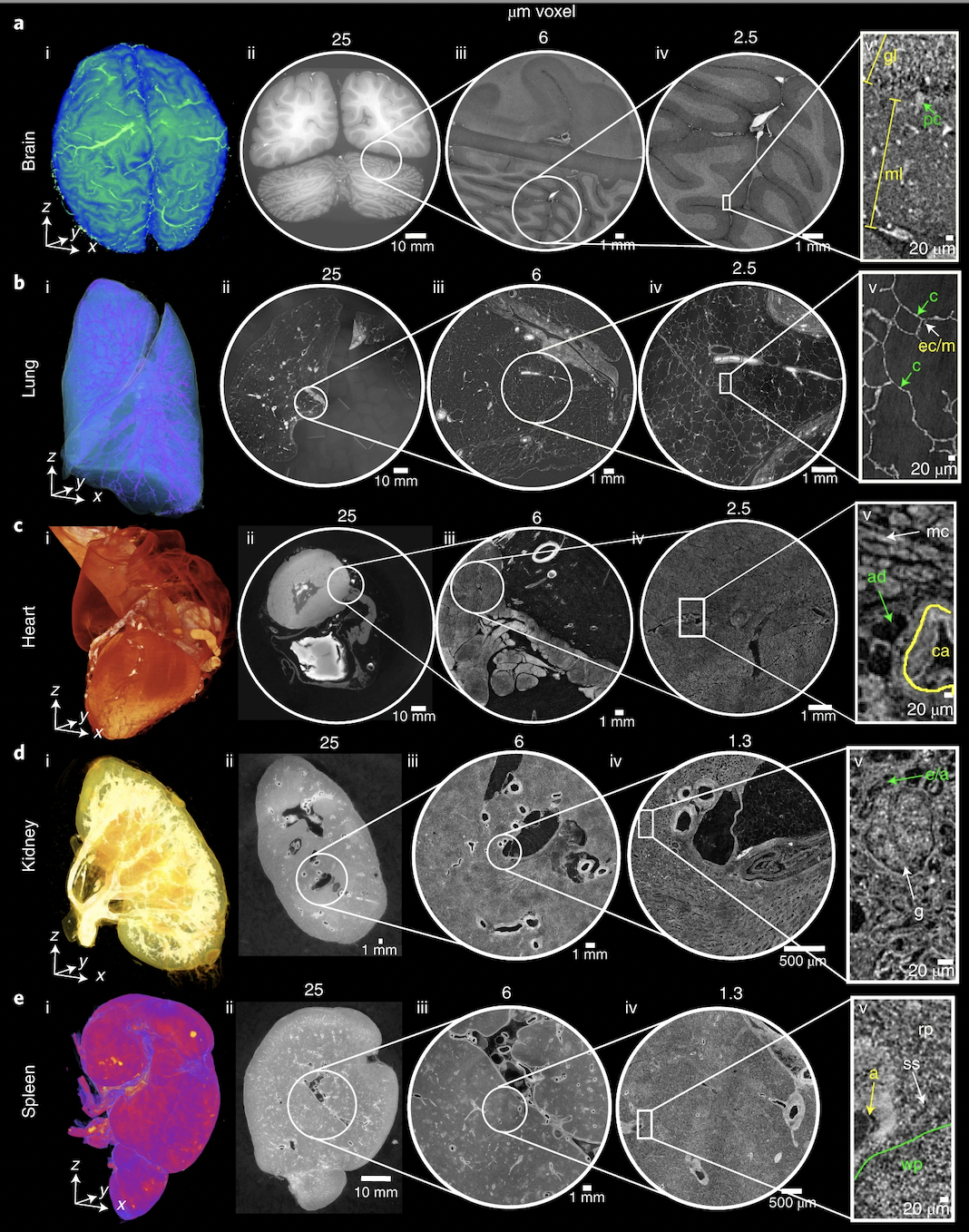}
    \caption{Figure after \cite{walsh2021imaging}, HiP-CT of brain (a), lung (b), heart (c) kidney (d) and spleen (e); for each organ, a 3D rendering (i) of the whole organ is shown using scans at 25$\mu$m per voxel. Subsequent 2D slices (ii–iv) show positions of the higher-resolution VOI relative to the previous scan. (v), Digital magnification of the highest-resolution image with annotations depicting characteristic structural features in the brain (ml, molecular layer; gl, granule cell layer; pc, Purkinje cell), in the lung (c, blood capillary; ec/m, epithelial cell or macrophage), in the heart (mc, myocardium; ca, coronary artery; ad, adipose tissue), in the kidney (e/a, efferent or afferent arteriole; g, glomerulus) and in the spleen (rp, red pulp; wp, white pulp; a, arteriole; ss, splenic sinus). All images are shown using 2× binning. }
  \label{fig:hipct}
\end{figure*}

A specific novelty of HiP-CT imaging as it relates to the vasculature, lies in its hierarchical nature. Whole human organs can be imaged with overview resolution of $\sim20\mu{m}/$voxel, then followed by high resolution (down to $\sim2\mu{m}/$voxel) regions of interest anywhere within the sample without sectioning (Figure \ref{fig:hipct}). This unique feature allows researchers to image the hierarchical structure of the vascular tree from the largest vessels down to near the capillary bed in intact human organs. This novel capability has already allowed unique insights into vascular abnormalities in COVID-19 lung lobes to be identified \cite{ackermann2020pulmonary,ackermann2022fatal,mentzer2022endothelialitis}, as well as vascular quantification across the whole organ \cite{rahmani2023micro}. Despite its advantages and suitability for vessel imaging, HiP-CT also poses many challenges for CNN-based methods of vascular segmentation:

\begin{itemize}
    \item {large data volume (routinely $\geq$ 1Tb) require cutting-edge hardware}
    \item { multi-scale vessels to segment}
    \item {lack of available ground truth training data due to the novelty of the method}
    \item {phase contrast generates fringes in intensity at vessel boundaries}
    \item{large vessels can collapse}
    \item {no targeted vascular staining to enhance contrast between vessel lumen and surrounding tissue}
    \item{continually developing technique with ongoing changes in image quality (e.g. improvement to resolution, SNR and CNR)  result in heterogeneous datasets}
\end{itemize}

In light of the opportunity and challenges that HiP-CT poses for vascular research, we perform a review of the literature on blood vessel segmentation with a particular focus on identifying current models that are most suited to the segmentation of blood vessels from HiP-CT imaging. We review segmentation on an organ-by-organ basis as the vascular and parenchymal tissue structures surrounding the vasculature differ vastly between different tissues. Our ultimate aim is to identify and then develop a model and training strategy which will provide a robust and generalisable method that can be easily adapted to segment multi-scale vascular structures across all organs imaged by HiP-CT.  

This paper is arranged as follows: In Section \ref{sec2}, deep learning models used for vessel segmentation, the datasets used in their training, and the evaluation metrics in the literature are briefly explained. In Section \ref{sec3}, the most prominent method in the literature is tested on HiP-CT kidney vascular data to establish a baseline of performance and highlight specific challenges in the adaptation of these models to HiP-CT data. In the final Section, the discussion and conclusions are presented.

\section{Vessel segmentation}\label{sec2}

\subsection{Modalities for 3D vascular imaging}

Whilst the ideal vessel segmentation algorithm or model is robust across a range of imaging modalities, in reality, the imaging modality will impose contrast and resolution limits that must be considered in segmentation approaches as well as imaging-specific artefacts (e.g anisotropic voxels).  
Here, we review some of the most common imaging modalites for vascular networks and contextualise the development of HiP-CT within this field. 
\par The most widely adopted procedure for visualising vessels in human organs \textit{in vivo} is angiography, i.e. where a contrast agent is injected into the vessels. The major angiographic modalities utilised in clinical practice are computed tomography angiography (CTA), magnetic resonance angiography (MRA), and digital subtraction angiography (DSA) also known as a conventional angiogram. 

MRA emerged as the predominant modality in vessel segmentation studies, as evidenced in Tables \ref{tab:brain}, \ref{tab:kidney}, \ref{tab:coronary}, and \ref{tab:lung}. Among the 31 deep learning-based vessel segmentation studies we reviewed across various organs, 9 employed MRA. Notably, for brain vessel segmentation, 9 out of the 12 studies utilised this modality. This is largely attributed to the fact that commonly available datasets for brain vasculature studies, such as PEGASUS \cite{mutke2014clinical}, MIDAS \cite{bullitt2005vessel}, SCAPIS \cite{bergstrom2015swedish}, and 1000PLUS \cite{hotter2009prospective}, utilise MRA.

MRA can be divided into three categories depending on the technique behind the image acquisition / generation: TOF (time-of-flight), PC (phase contrast), and CE (contrast-enhanced). Among those,  TOF-MRA is the most frequently used non-contrast bright-blood technique for imaging the human vascular system \cite{forkert2009automatic,bollmann2022imaging}. As suggested by its name, TOF MRA relies on a principle known as flow-related enhancement, which happens when completely magnetised blood flows into a slab of magnetically saturated tissue whose signal has been muted by repeated RF-pulses \cite{baghaie2018curvelet}. However, the long acquisition time and high operational costs make it difficult to employ for cerebral artery visualisation \cite{klimont2020deep,katz1995circle}. Furthermore, MRA has been shown to overestimate stenosis compared to other modalities \cite{buerke2009dual}. 

CTA, on the other hand, is faster, takes a few minutes to complete, and is more accurate than MRA. However, unlike MRA, all CTAs need the administration of an IV contrast agent and involve radiation exposure, making it less safe.

DSA has been utilised as the gold standard for imaging vessels; nonetheless, this invasive and labor-intensive method is rather costly and comes with discomforts and possible dangers\cite{willinsky2003neurologic,kaufmann2007complications}. 
For the \textit{in vivo} methods described above the resolution of imaging is on the order of 0.5-2mm \cite{bollmann2022imaging,willinsky2003neurologic},  with the state-of-the-art TOF techniques able to capture at most (300$\mu{m}$  - 500$\mu{m}$).
Capturing the microvasculature (1-100$\mu{m}$ ) is only possible \textit{in vivo} at very shallow depths using e.g. contrasted ultrasound\cite{kierski2020perspectives}, or must be performed \textit{ex vivo} on extracted tissue using optical or X-ray methods \cite{jafree2019spatiotemporal,reichardt20213d}. Even within the \textit{ex vivo} techniques, there are only two, Light-sheet imaging and HiP-CT capable of imaging intact human organ vascular networks that bridge the length scales from large arteries and veins down to the capillary bed \cite{zhao2020cellular,walsh2021imaging}. Of the two techniques, HiP-CT can perform multi-scale imaging far more quickly than light sheet imaging as HiP-CT does not rely on staining of the vasculature shortening sample preparation, \cite{rahmani2023micro}, and imaging itself of a whole organ at 20$\mu{m}$ /voxel can be performed in as little as 1 hr \cite{rahmani2023micro}. Whilst the acquisition of these large vascular network datasets is becoming increasingly fast, automated segmentation has also been a rapidly developing area of research \cite{wagner20213d}.

\begin{table}
\centering
\caption{Review of the papers applying deep learning for brain vessel segmentation. CNN: Convolutional neural network; GAN: Generative adversarial network; cGAN: Conditional GAN; DD-Net: Dense-dilated neural network; LSTM: Long short-term memory; U-Net MSS: Multi-scale supervised U-Net; TOF MRA: Time of flight magnetic resonance angiography; DSA: Digital subtraction angiography; $\mu$CTA: Micro-computed tomography angiography; 3DRA: 3D rotational angiographies; DSC: Dice similarity coefficient; HD: Hausdorff distance; AVD: Average distance; Prec: Precision; MHD: Mahalanobis distance; IOU: Intersection over union; PPV: Positive predictive value; CAL: Connectivity-area-length.  }
\label{tab:brain}
\resizebox{\textwidth}{!}{%
\begin{tabular}{@{}cccccccc@{}}
\toprule
\textbf{References}                                             & \textbf{Modality}                                    & \textbf{Data source}                                                            & \textbf{No. of subj.} & \textbf{ML model}                                                                           & \textbf{Input}                                               & \textbf{DSC}                                                                           & \textbf{Add'l perf. metrics}                                                            \\ \midrule
Livne et al. \cite{livne2019u}                 & TOF MRA                                                     & \begin{tabular}[c]{@{}c@{}}Private \\ (PEGASUS)\end{tabular}                    & 66                       & U-Net                                                                                                     & 2D patches                                                   & 0.88                                                                                   & \begin{tabular}[c]{@{}c@{}}95HD = $\sim$47 voxels \\ and an \\ AVD = $\sim$0.4 voxels\end{tabular} \\
Phellan et al. \cite{phellan2017vascular}      & TOF MRA                                                     & Private                                                                         & 5                        & 2D CNN                                                                                                    & 2D patches                                                   & \begin{tabular}[c]{@{}c@{}}between 0.764 \\ and 0.786\end{tabular}                     & -                                                                                                  \\
Hilbert et al. \cite{hilbert2020brave}         & TOF MRA                                                     & \begin{tabular}[c]{@{}c@{}}Private \\ (PEGASUS+\\ 7UP+\\ 1000Plus)\end{tabular} & 264                      & \begin{tabular}[c]{@{}c@{}}3D CNN \\ (BRAVE-NET)\end{tabular}                                             & 3D patches                                                   & 0.931                                                                                  & \begin{tabular}[c]{@{}c@{}}95HD = 29.153, \\ and AVD = 0.165\end{tabular}                          \\
Patel et al. \cite{patel2020multi}             & DSA                                                         & Private                                                                         & 100                      & \begin{tabular}[c]{@{}c@{}}3D CNN (DeepMedic) \\ and 3D U-Net\end{tabular}                                & 3D patches                                                   & \begin{tabular}[c]{@{}c@{}}0.94±0.02 \\ and 0.92±0.02 \\ respectively\end{tabular}     & \begin{tabular}[c]{@{}c@{}}CAL= 0.84±0.07 \\ and 0.79±0.06 \\ respectively\end{tabular}            \\
Tetteh et al. \cite{tetteh2020deepvesselnet}   & \begin{tabular}[c]{@{}c@{}}TOF MRA \\ and $\mu$CTA\end{tabular} & \begin{tabular}[c]{@{}c@{}}Publicly \\ available\end{tabular}                   & Synthetic data           & \begin{tabular}[c]{@{}c@{}}NN with 2D \\ orthogonal \\ cross-hair filters \\ (DeepVesselNet)\end{tabular} & 3D volumes                                                   & 0.79                                                                                   & \begin{tabular}[c]{@{}c@{}}Prec=0.77 \\ and Recall=0.82\end{tabular}                               \\
Garcia et al. \cite{garcia2022deep}            & 3DRA                                                        & Private                                                                         & 5                        & \begin{tabular}[c]{@{}c@{}}3DUNet-based \\ architectures\end{tabular}                                     & 3D patches                                                   & 0.80 ± 0.06                                                                            & \begin{tabular}[c]{@{}c@{}}Prec=0.75 ± 0.10\\  and Recall=0.90 ± 0.07\end{tabular}                 \\
Vos et al. \cite{de2021automatic}              & TOF-MRA                                                     & Private                                                                         & 69                       & \begin{tabular}[c]{@{}c@{}}2D and \\ 3D U-Net\end{tabular}                                                & \begin{tabular}[c]{@{}c@{}}2D and \\ 3D patches\end{tabular} & \begin{tabular}[c]{@{}c@{}}0.74 ± 0.17 \\ and 0.72 ± 0.15 \\ respectively\end{tabular} & \begin{tabular}[c]{@{}c@{}}MHD=47.6 ± 40.4 \\ and 5 81.3 ± 57.0 \\ respectively\end{tabular}       \\
Chatterjee et al. \cite{chatterjee2020ds6}     & TOF MRI                                                     & \begin{tabular}[c]{@{}c@{}}Not explicity \\ stated\end{tabular}                 & 11                       & U-Net MSS                                                                                                 & 3D patches                                                   & 0.79 ± 0.091                                                                           & IOU=65.89±1.25                                                                                     \\
Zhang and Chen \cite{zhang2019ddnet}           & TOF-MRA                                                     & \begin{tabular}[c]{@{}c@{}}Not explicity \\ stated\end{tabular}                 & 42                       & DD-Net                                                                                                    & 3D patches                                                   & 0.67                                                                                   & \begin{tabular}[c]{@{}c@{}}Sensitivity=67.86 \\ and IOU=33.66\end{tabular}                         \\
B.Zhang et al. \cite{zhang2020cerebrovascular} & TOF-MRA                                                     & \begin{tabular}[c]{@{}c@{}}Publicly available \\ (MIDAS)\end{tabular}           & 109                      & DD-CNN                                                                                                    & 3D patches                                                   & 0.93                                                                                   & \begin{tabular}[c]{@{}c@{}}PPV=96.4730, \\ Sensitivity=90.1443, \\ and Acc=99.9463\end{tabular}    \\
Lee et al. \cite{lee2021spider}                & MRA                                                         & \begin{tabular}[c]{@{}c@{}}Private \\ (SNUBH)\end{tabular}                      & 26                       & \begin{tabular}[c]{@{}c@{}}2D U-Net with LSTM \\ (Spider U-Net)\end{tabular}                              & `2D images                                                   & 0.793                                                                                  & IOU=74.3                                                                                           \\
Quintana et al. \cite{quintana2022dual}        & TOF-MRA                                                     & \begin{tabular}[c]{@{}c@{}}Private \\ (UNAM)\end{tabular}                       & 4                        & Dual U-Net-Based cGAN                                                                                     & 2D images                                                    & 0.872                                                                                  & Prec=0.895                                                                                         \\ \bottomrule
\end{tabular}%
}
\end{table}

\begin{table}
\centering
\caption{Review of the papers applying deep learning for kidney vessel segmentation. CNN: Convolutional neural network; GAN: Generative adversarial network; WSI: Whole slide imaging; DPA: Deep priori anatomy; MCD: Mean centerline distance; EnMcGAN: Ensemble multi-condition GAN; DUP-Net: Double UPoolFormer networks; DSC: Dice similarity coefficient. }
\label{tab:kidney}
\resizebox{\textwidth}{!}{%
\begin{tabular}{@{}cccccccc@{}}
\toprule
\textbf{References}                             & \textbf{Modality} & \textbf{Data source} & \textbf{No. of subj.} & \textbf{ML model}                                                         & \textbf{Input} & \textbf{DSC}                                                                  & \textbf{Add'l perf. metrics}                                                   \\ \midrule
Karpinski et al. \cite{salvi2020karpinski}     & WSI               & Publicly available   & 35                    & \begin{tabular}[c]{@{}c@{}}2D UNet with \\ Resnet34 backbone\end{tabular} & 2D images      & -                                                                             & Acc: 0.893                                                                     \\
He et al. \cite{he2019dpa}                     & CT                & Private              & 170                   & DPA-DenseBiasNet                                                          & 3D volumes     & 0.861                                                                         & MCD:1.976                                                                      \\
Taha et al. \cite{taha2018kid}                 & CT                & Private              & 99                    & Kid-Net (a 3D CNN)                                                        & 3D patches     & -                                                                             & \begin{tabular}[c]{@{}c@{}}F1 score: 0.72 (artery);\\ 0.67 (vein)\end{tabular} \\
He at al. \cite{he2021enmcgan} & CTA               & Private              & 122                   & EnMcGAN                                                                   & 3D patches     & \begin{tabular}[c]{@{}c@{}}0.89±0.6 (artery);\\ 0.77±0.12 (vein)\end{tabular} & -                                                                              \\
Zhang et al. \cite{zhang2022novel}             & CT                & Publicly available   & 392                   & DPA-DenseBiasNet                                                          & 2D images      & 0.884                                                                         & -                                                                              \\
Xu et al. \cite{xu2023extremely}               & micro-CT        & Private              & 8                     & CycleGAN                                                                  & 3D patches     & 0.768 ± 0.3                                                                   & Acc: 0.992                                                                     \\
Li et al. \cite{li2023dup}     & CT                & Publicly available   & 35                    & DUP-Net                                                                   & 3D patches     & 0.883                                                                         & \begin{tabular}[c]{@{}c@{}}Precision:0.911;\\ Recall:0.858\end{tabular}        \\ \bottomrule
\end{tabular}%
}
\end{table}

\begin{table}
\centering
\caption{Review of the papers applying deep learning for coronary vessel segmentation. DSC: Dice similarity coefficient; Prec: Precision; Sen: Sensitivity; Spec: Specificity; GBDT: Gradient boosting decision tree; CCTA:Coronary computed tomographic angiography; XCA: X-ray coronary angiography; GCN: Graph convolutional networks; MSD: Mean surface
distance; AUC: Area under the receiver operating characteristic curve; ROI: Region of interest; 3D FFR U-Net: 3D feature fusion and rectification U-Net}
\label{tab:coronary}
\resizebox{\textwidth}{!}{%
\begin{tabular}{@{}cccccccc@{}}
\toprule
\textbf{References}                                         & \textbf{Modality} & \textbf{Data source}                                          & \textbf{No. of subj.} & \textbf{ML model}   & \textbf{Input} & \textbf{DSC}  & \textbf{Add'l perf. metrics}                                                  \\ \midrule
Dong et al. \cite{dong2022novel}           & CCTA              & Private                                                       & 338                   & Di-Vnet             & 3D patches     & 0.902         & \begin{tabular}[c]{@{}c@{}}Prec=0.921,\\ Recall=0.97\end{tabular}             \\
Gao et al. \cite{gao2022vessel}            & XCA               & Private                                                       & 130                   & GBDT                & 2D images      & -             & \begin{tabular}[c]{@{}c@{}}F1=0.874,\\ Sen= 0.902,\\  Spec=0.992\end{tabular} \\
Wolterink et al .\cite{wolterink2019graph} & CCTA              & \begin{tabular}[c]{@{}c@{}}Publicly \\ available\end{tabular} & 18                    & GCN                 & 2D images      & 0.74          & MSD=0.25mm                                                                    \\
Li et al. \cite{li2022automatic}           & CCTA              & Private                                                       & 243                   & 2D U-Net with 3DNet & 2D images      & 0.771 ± 0.021 & AUC=0.737                                                                     \\
Song et al. \cite{song2022automatic}       & CCTA              & Private                                                       & 68                    & 3D FFR U-Net             & 3D patches     & 0.816         & \begin{tabular}[c]{@{}c@{}}Prec=0.77,\\ Recall=0.87\end{tabular}              \\
Zeng et al. \cite{zeng2023imagecas}        & CCTA              & \begin{tabular}[c]{@{}c@{}}Publicly \\ available\end{tabular} & 1000                  & 3D-UNet             & 3D volumes     & 0.82          & -                                                                             \\ \bottomrule
\end{tabular}%
}
\end{table}

\begin{table}
\centering
\caption{Review of the papers applying deep learning for pulmonary vessel segmentation. CNN: Convolutional neural network; 
CT: computed tomography; CTA: computed tomography angiography; MSI-U-Net: Multi-scale interactive U-Net; OR: Over segmentation rate; AUROC: Area under the receiver operating characteristic curve; Sen: Sensitivity; Spec: Specificity; Prec: Precision. }
\label{tab:lung}
\resizebox{\textwidth}{!}{%
\begin{tabular}{@{}cccccccc@{}}
\toprule
\textbf{References}                                           & \textbf{Modality} & \textbf{Data source}                                         & \textbf{No. of subj.} & \textbf{ML model}                                             & \textbf{Input}                                                & \textbf{DSC}                                                             & \textbf{Add'l perf. metrics}                                       \\ \midrule
Tan et al. \cite{tan2021automated}           & CT and CTA        & \begin{tabular}[c]{@{}c@{}}Publicly\\ available\end{tabular} & 16                    & \begin{tabular}[c]{@{}c@{}}2D-3D U-Net\\ nnU-net\end{tabular} & \begin{tabular}[c]{@{}c@{}}2D images\\ 3D volume\end{tabular} & \begin{tabular}[c]{@{}c@{}}0.786,\\ 0.797,\\ 0.77\\ (on CT)\end{tabular} & \begin{tabular}[c]{@{}c@{}}OR=0.281,\\ 0.285,\\ 0.304\end{tabular} \\
Nam et al. \cite{nam2021automatic}           & CTA               & Private                                                      & 104                   & 3D U-Net                                                      & 3D patches                                                    & 0.915±0.31                                                               & AUROC=0.995                                                        \\
Guo et al. \cite{guo2021comparison}          & CT                & Private                                                      & 50                    & 3D CNN                                                        & 3D patches                                                    & 0.943                                                                    & -                                                                  \\
Xu et al. \cite{xu2018pulmonary}             & CT                & Private                                                      &                       & 2D CNN                                                        &                                                               & -                                                                        & -                                                                  \\
Nardelli et al. \cite{nardelli2018pulmonary} & CT                & \begin{tabular}[c]{@{}c@{}}Publicly\\ available\end{tabular} & 55                    & 3D CNN                                                        & \begin{tabular}[c]{@{}c@{}}2D and \\ 3D patches\end{tabular}  & -                                                                        & \begin{tabular}[c]{@{}c@{}}Sen=0.93\\ Prec=0.83\end{tabular}       \\
Wu et al. \cite{wu2023multi}                 & CT                & Private                                                      & 143                   & MSI-U-Net                                                     & 3D volume                                                     & 0.7168                                                                   & \begin{tabular}[c]{@{}c@{}}Sen=0.7234,\\ Prec=0.7893\end{tabular}  \\ \bottomrule
\end{tabular}%
}
\end{table}

Before the arrival of complex machine learning methods, vessel segmentation relied heavily on traditional techniques such as kernel-based methods, tracking methods, mathematical morphology-based methods, and model-based methods \cite{taha2015metrics}. In kernel-based methods, edge detection techniques, like the Canny or Sobel operators, identify vessel boundaries using gradient information \cite{chatterjee2021retinal}. More sophisticated approaches,  such as Hessian-based methods can be categorised under model-based methods \cite{foruzan2012hessian}. These methods utilise second-order intensity derivatives to capture vessel-like structures, and level-set methods evolve contours within the image domain to detect vessels. Active contours, or snakes, are tracking methods dynamically adjusted to capture vessel boundaries, and mathematical tools like B-splines offer precise modeling of vessel structures \cite{zhao2015automated,shang2010vascular}. 

These traditional methods provided a foundation that paved the way for machine learning in the realm of medical image analysis.

With the rapid advancement of computational capabilities, a paradigm shift has taken place in the realm of image processing and machine learning: the rise of deep neural networks (NNs). These NNs, employing multi-layered architectures, extracted patterns within data through multiple levels of abstraction, and transforming the way images are understood and processed.

Among various NN architectures, CNNs have stood out as highly effective for image analysis tasks. They possess the capability to learn image features in a hierarchical manner, progressing from simple to complex, thereby eliminating the need for manual feature extraction. This inherent feature has made CNNs especially popular in the field of medical image analysis.

Recently, the landscape of computer vision has seen another transformative shift with the emergence of Transformers \cite{vaswani2017attention}. Initially successful in natural language tasks, Transformers have begun to find application in various computer vision challenges, prompting researchers to reevaluate the dominant role of CNNs. These advancements in computer vision have also generated significant interest in the medical imaging domain \cite{shamshad2023transformers}.
Transformers, with their ability to capture a more extensive global context compared to CNNs have now become a subject of considerable attention in the medical imaging community \cite{al2023vision,he2023transformers}.

\subsection{Evaluation metrics}

\begin{figure*}[!ht]
\centering
  \includegraphics[width=0.8\textwidth]{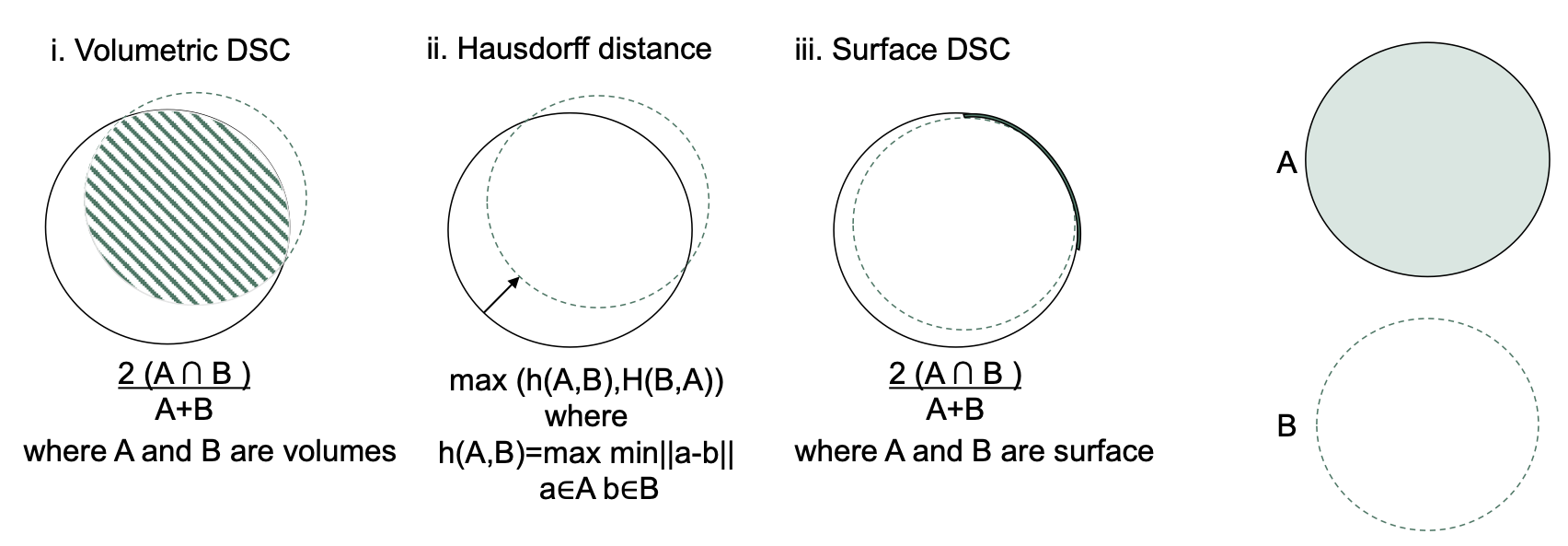}
    \caption{Illustration for calculation of volumetric Dice score coefficient (DSC), Hausdorff distance, and surface DSC. The solid line represents the ground truth contour, whereas the dashed line is the prediction. i.  A Volumetric DSC, defined as the union of two volumes (green volume region) normalised by the mean of the two volumes. ii. Hausdorff distance, defined as the maximum nearest neighbor Euclidean distance (green arrow). iii. Surface DSC, is defined as the union of two contours (yellow contour region) normalised by the mean surface of the two contours. - after \cite{vaassen2020evaluation}}
  \label{fig:eval_metics}
\end{figure*}

In order to evaluate any automated vessel segmentation model or algorithm, a quantitative assessment of a segmentation task can be done through various performance metrics. Depending on the theory behind them, these metrics can be divided into overlap-based, volume-based, pair-counting-based, information-theoretic-based, probabilistic-based, and spatial distance-based measures \cite{taha2015metrics}. These are summarised in Figure \ref{fig:eval_metics}

When assessing the performance of segmentation algorithms in comparison to ground truth, a contingency table is often employed, featuring True Positive (TP), True Negative (TN), False Negative (FN), and False Positive (FP) values. In this context, positive and negative denote pixels that, according to the ground truth segmentation, are attributed to vessels and background respectively.

A review of the literature reveals varied metric reporting across studies (as evidenced in Tables \ref{tab:brain}, \ref{tab:kidney}, \ref{tab:coronary}, and \ref{tab:lung}). This section delves into the most frequently used performance metrics, elucidating their underlying theory, strengths, and weaknesses.

\textbf{Overlap-Based Measures} Overlap-based measures are used to evaluate the similarity between the segmented result and the ground truth based on their overlapping regions. Such measures are the most widely utilised metrics with all studies we reviewed reporting at least one overlap based metric. However they have some limitations when applied to vascular segmentation which will be discussed further.
The Dice Similarity Coefficient (DSC) is one of the most common overlap-based metrics that calculates the spatial overlap of two segmentations 26 of the 31 studies reviewed here utilise DSC. Given its properties, DSC is particularly used in vascular segmentation where the vasculature accounts for a small fraction (e.g. around 3\% in the brain) of the organ volume, leading to unbalanced data challenges \cite{nicholson2001diffusion}.

\begin{equation}
    Dice = \frac{2 \cdot TP}{FP+FN+(2 \cdot TP)} \mbox{.}
\end{equation}

Sensitivity, also known as TP Rate or Recall, is another overlap-based metric. It measures how well a model identifies TPs (e.g., correctly segmented vessels). It provides insight into how much of the ground truth overlaps with the predicted positive segment. Specificity, on the other hand, measures the proportion of TNs (e.g., correctly identified non-vessel areas) that are correctly detected. It reflects on how much of the ground truth's negative area overlaps with the predicted negative segment. These two measures are defined as follows:

\begin{equation}
Sensitivity=\frac{TP}{TP+FN} \mbox{,}
\end{equation}

\begin{equation}
Specificity=\frac{TN}{FP+TN} \mbox{.}
\end{equation}

Another related measure is precision, often referred to as the positive predictive value (PPV). While not frequently utilised in medical image validation, it plays a role in determining the F-Measure. This is defined as

\begin{equation}
Precision=\frac{TP}{TP+FP} \mbox{,}
\end{equation}

whereas F-Measure, also known as F1 score is the harmonic mean of precision and recall. 

In 2020, Shit et al. developed centerline Dice (\textit{clDice}), a unique topology-preserving loss function for tubular structure segmentation, as another DSC modification \cite{shit2021cldice}. Their work highlighted that DSC metric does not necessarily assess the connectedness of a segmentation because Dice does not equally weight tubular structures with large, medium, and small radii. As connectivity is a crucial biological feature of vascular networks, bespoke metrics should be used to evaluate vascular segmentations. The authors demonstrated that training on a globally averaged loss causes a considerable bias towards the volumetric segmentation of big arteries in real vascular datasets compared to arterioles and capillaries (radius ranges e.g., 30 $\mu$m for arterioles and 5 $\mu$m for capillaries). To enable topology preservation, centerline Dice is formulated as follows:

\begin{equation}
T_{\text{prec}} = \frac{\mid S_{P} \cap V_{L} \mid}{\mid S_{P}\mid} \mbox{,}
\end{equation}

\begin{equation}
T_{\text{sens}} = \frac{\mid S_{L} \cap V_{P} \mid}{\mid S_{L} \mid} \mbox{,}
\end{equation}

\begin{equation}
clDice = \frac{2 \times T_{\text{prec}} \times T_{\text{sens}}}{T_{\text{prec}} + T_{\text{sens}}} \mbox{,}
\end{equation}

where the ground truth mask is $V_{L}$, the predicted segmentation mask is $V_{P}$ and
skeletons are $S_{P}$ and $S_{L}$.

\textbf{Volume-Based Measures} These metrics focus on the volume or size of the segmented structures. 
As the name implies, Volume Difference (VD) is a typical volume-based metric that is based on the absolute difference in volumes of the segmented structure and the ground truth, usually normalised by the ground truth volume. Contrary to a simple volumetric subtraction, VD accounts for the volumetric overlap by considering voxels exclusive to each mask as well as those shared. This ensures that VD accurately reflects both the volume and the spatial agreement between the two segmentations. VD is formulated as follows:

\begin{equation}
VD = \frac{\lvert V_p - V_g \rvert}{V_g} \mbox{,}
\end{equation}

where $V_p$ is the volume of the predicted segmentation and $V_g$ is the volume of the ground truth.

\textbf{Pair-Counting-Based Measures} The pair counting based measures are calculated based on the correspondence between object pairs in the segmentation and the ground truth. One such metric is Adjusted Rand Index (ARI) and it is calculated as follows:

\begin{equation}
\text{ARI} = \frac{\text{RI} - \text{Expected\_RI}}{\text{Max\_RI} - \text{Expected\_RI}} \mbox{,}
\end{equation}

where RI is the Rand Index and is calculated as:

\begin{equation}
\text{RI} = \frac{\text{a} + \text{d}}{\text{a} + \text{b} + \text{c} + \text{d}} \mbox{,}
\end{equation}

where a is the number of pairs of objects that are in the same group in both the predicted and the ground truth and corresponds to the TPs; d is the number of pairs of objects that are in different groups in both and corresponds to the TNs; and b and c correspond to FPs and FNs, respectively.


\textbf{Information-Theoretic-Based Measures} As the name implies, information-theoretic-based measures use information theory concepts to estimate the quality and performance of the segmentation. For instance, one such measure called Mutual Information (MI) is calculated based on the shared information between the segmented result and the ground truth.

For two discrete random variables X (segmentation result) and Y (ground truth), the MI is defined as:

\begin{equation}
    MI(X,Y) = \sum_{x \in X} \sum_{y \in Y} p(x,y) \log \left( \frac{p(x,y)}{p(x)p(y)} \right) \mbox{,}
\end{equation}

where 
 
\begin{itemize}
    \item \( p(x,y) \) is the joint probability distribution function of \(X\) and \(Y\).
    \item \( p(x) \) is the marginal probability distribution function of \(X\).
    \item \( p(y) \) is the marginal probability distribution function of \(Y\).
\end{itemize}

\(X\) might represent the predicted segmentation where \\
a specific value \(x\) taken by \(X\) could be either ``object'' (e.g., a vessel) or ``background.'' and \(Y\) represents the ground truth (actual segmentation) where \\
a specific value \(y\) taken by \(Y\) could similarly be either ``object'' or ``background.''

In that context:

\begin{itemize}
    \item \( p(x=\text{object},y=\text{object}) \) corresponds to the probability of a TP.
    \item \( p(x=\text{background},y=\text{background}) \) corresponds to the probability of a TN.
    \item \( p(x=\text{object},y=\text{background}) \) corresponds to the probability of a FP.
    \item \( p(x=\text{background},y=\text{object}) \) corresponds to the probability of a FN.
\end{itemize}

\textbf{Probabilistic-Based Measures}

Probabilistic-based measures evaluate the performance of segmentation and classification algorithms based on predicted probabilities rather than strict binary or discrete decisions. In binary classification, for example, a probabilistic model may assign a probability indicating the likelihood that an item belongs to the positive class rather than simply categorizing it as positive or negative. 

The ROC curve (Receiver Operating Characteristic) stands as a prime example of a probabilistic-based measure. It graphs the TP rate (sensitivity) against the TN rate (specificity) over a spectrum of decision thresholds. This curve offers a holistic perspective on a model's capability to distinguish between classes.

The Area Under the Curve (AUC) encapsulates the model's overall discriminative prowess between positive and negative classes. From a mathematical standpoint, the AUC represents the integral of the ROC curve:

\begin{equation}
\text{AUC} = \int_0^1 \text{TPR}(t)dt \mbox{,}
\end{equation}

where \(\text{TPR}(t)\) is the TP rate at a given threshold \(t\).
This metric or variations appear in 3 of the 31 models reviewed in Tables \ref{tab:brain}, \ref{tab:kidney}, \ref{tab:coronary}, and \ref{tab:lung}).

\textbf{Spatial Distance-Based Measures}

Spatial distance-based measures focus on the spatial discrepancies between the segmented structures and the ground truth. For instance, the 95th Percentile Hausdorff Distance (95HD) assesses the greatest distance between a point in the true segmentation and the nearest point in the segmented result. More formally, if we define \( h(A, B) \) as the set of all distances from points in set \( A \) to their nearest points in set \( B \), then:

\begin{equation}
    \text{95HD}(A, B) = \text{percentile}(h(A, B), 95) \mbox{.}
\end{equation}

The Average Volumetric Distance (AVD), on the other hand, is typically computed as the absolute difference in volumes between the predicted segmentation \( V_p \) and the ground truth \( V_{gt} \), normalised by the volume of the ground truth. Mathematically, it can be expressed as:

\begin{equation}
\text{AVD} = \frac{\mid V_{p} - V_{gt} \mid}{V_{gt}} \times 100 \mbox{.}
\end{equation}

Boundary-based techniques are another important subcategory of distance-based measures \cite{maier2022metrics}. Normalised surface Dice, mean average surface distance, and average symmetric surface distance are three key metrics, especially in the field of vessel segmentation. Surface dice is a measurement to evaluate the similarity between the segmented surface and a ground truth surface. Normalised surface Dice takes this a step further by normalising the measurement and making it less sensitive to size differences between the predicted segmentation and the ground truth. This is particularly useful for segmenting smaller structures, or size variability between subjects. Surface distance is another measurement that indicates the shortest distance of each point on the surface of the segmented structure to the surface of the ground truth and is calculated as follows:

\begin{equation}
\text{ASSD}(A,B) = 0.5\left( \frac{1}{\mid A \mid} \sum_{a \in A} \min_{b \in B} \text{dist}(a,b) + \frac{1}{\mid B \mid} \sum_{b \in B} \min_{a \in A} \text{dist}(b,a) \right) \mbox{,}
\end{equation}

where \text{dist}(a, b) is the Euclidean distance between points $a$ and $b$, and 
$\mid A \mid$ and $\mid B \mid$ are the numbers of surface points in A and B respectively.

Of the 31 paper reviewed 5 use a spatial distance metric. 

The use of the appropriate assessment metric is crucial, as each measurement has specific biases dependent on the characteristics of the segmented structures. Whilst our review of literature shows that DCS or other generalised overlap measures are still the most frequently used metrics, more specialised vascular specific metrics are available and are becoming more widely implemented. The evaluation metric can strongly influence the choice of the final optimal model and hence should be chosen in accordance with the segmentation task's nature and taking into consideration what downstream analyses are to be conducted on the segmentations, or the biological implications of a particular metric. For example for vascular structures, quantitative analyses of network topology for flow simulations are common end goals for segmentation \cite{rahmani2023micro,sweeney2018role}. In such cases meshing or skeletonising are common steps after segmentation.  Meshing or skeletonisation algorithms are often highly sensitive to holes or cavities in the segmentations, or to breaks in connectivity in vascular components\cite{zhang2022techniques}. Thus, evaluation methods highly sensitive for connectivity breaks (e.g. clDice) should be implemented in these cases.





\subsection{Deep learning models for lung, kidney, cardiovascular, and cerebral vessel segmentation}\label{sec2.2}

In this section, we delve into the latest deep-learning (DL) strategies used for blood vessel segmentation. We've grouped the DL-based techniques for segmenting vessels under three primary network structures: Convolutional Neural Networks, Generative Adversarial Networks, and Vision Transformers.

\subsubsection{Convolutional neural network based models (CNNs)}

Traditional CNNs have been foundational in vessel segmentation before the widespread adoption of architectures like U-Net. These neural networks, based on fundamental concepts of learning features and building hierarchical representations, leverage the presence of spatial structures and patterns inherent in image data. Unlike the Fully Convolutional Networks (FCN) that perform dense predictions, traditional CNNs often operate on patches or regions, aiming at classifying central pixels or aggregates. 

Yao et al. used a 2D CNN architecture to extract blood vessels from fundus images \cite{yao2016convolutional}. The output of their CNN architecture is the confidence level of each pixel being blood vessels. In 2018, Tetteh et al. developed an architecture called DeepVesselNet. Their design leverages 2-D orthogonal cross-hair filters using 3-D context information while minimizing computational demands.  Addinionally, they introduce a class balancing cross-entropy loss function with false positive rate correction specifically tailored to address the prevalent issues of significant class imbalances and high false positive rates observed with traditional loss functions \cite{tetteh2020deepvesselnet}.

Cervantes-Sanchez et al. trained a multilayer perceptron with X-ray Coronary Angiography (XCA) images enhanced by using Gaussian filters in the spatial domain and Gabor filters in the frequency domain for segmentation of coronary arteries in X-ray angiograms \cite{cervantes2019automatic}. Nasr-Esfahani et al. presented a multi-stage model where a patch around each pixel is fed
into a trained CNN to determine whether the pixel is of vessel or background regions.\cite{nasr2018segmentation}. Samuel and Veeramalai proposed a two-stage vessel extraction framework to learn well-defined vessel features from global features (learned by the pre-trained VGG-16 base network) using the Vessel Specific Convolutional (VSC) blocks, Skip chain Convolutional (SC) layers, and feature map summations \cite{samuel2021vssc}. 
In 2021, Iyer et al. developed a new CNN for angiographic segmentation: AngioNet, which combines an Angiographic Processing Network (APN) with Deeplabv3+ because of its ability to approximate more complex functions. The APN was tailored to tackle several challenges specific to angiographic segmentation, including low-contrast images and overlapping bony structures\cite{iyer2021angionet}. 

\begin{figure*}[!ht]
\centering
  \includegraphics[width=1\textwidth]{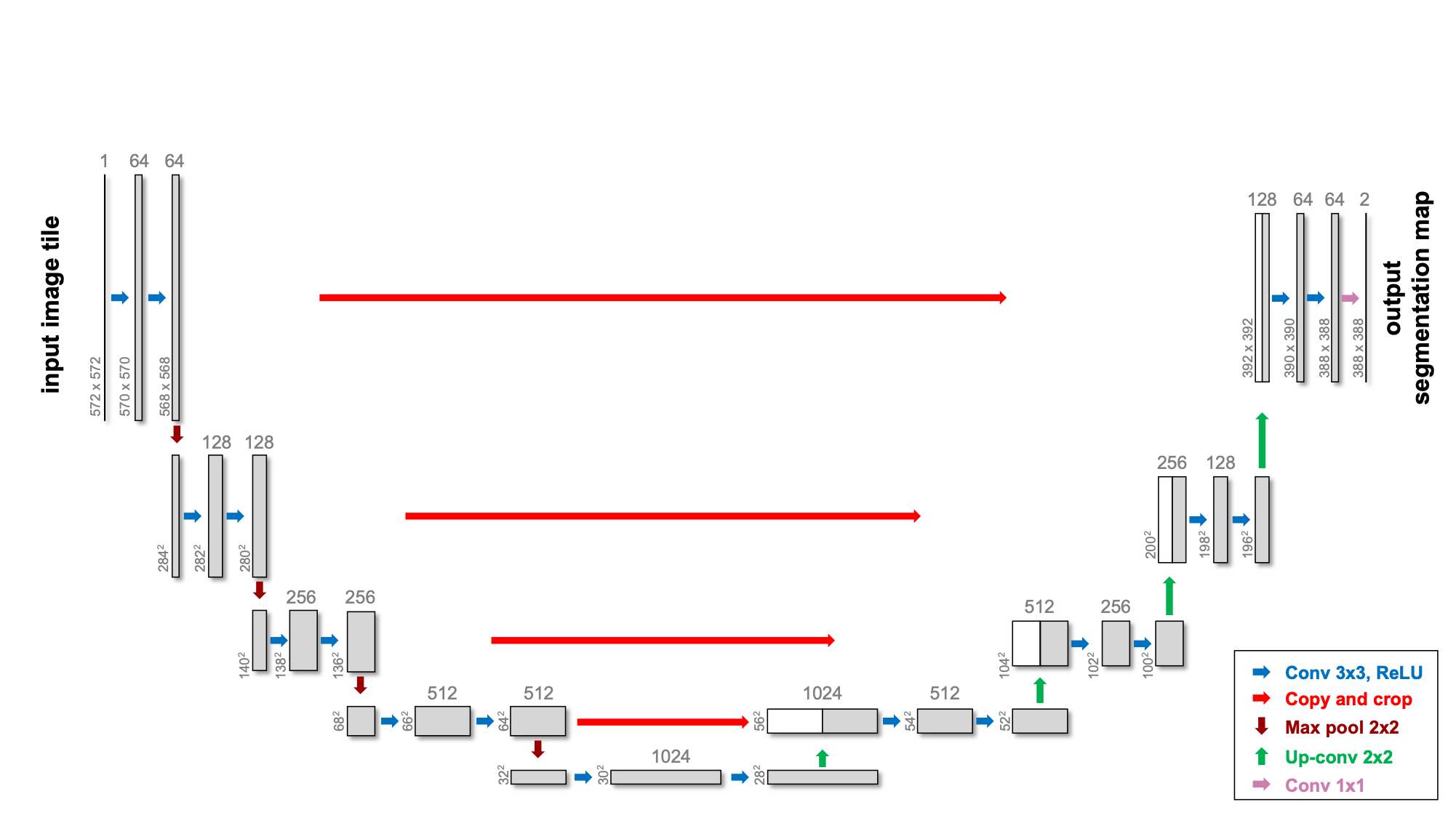}
    \caption{A sample U-Net architecture used in medical image segmentation tasks - after \cite{ronneberger2015u}}
  \label{fig:unet}
\end{figure*}

\paragraph{Fully convolutional neural network (FCN) based models: U-Net and its variants}

Fully Convolutional Networks (FCNs) are a subtype of CNNs specialised in semantic segmentation tasks. Deep learning techniques have revolutionised medical image segmentation, achieving precise pixel-level classification with the help of FCN \cite{long2015fully}. FCNs are designed to produce an output of the same size as the input (i.e., a pixel-wise map). They replace the dense layers with convolutional layers. Expanding upon FCN's foundation, many researchers have pioneered advanced 2D fully convolutional neural networks, including U-Net, SegNet, DeepLab, and PSPNet \cite{ronneberger2015u,badrinarayanan2017segnet, chen2017deeplab, zhao2017pyramid}. Khan et al., presented a fully convolutional network called RC-Net for the purpose of retinal image segmentation. The network itself is relatively small and the number of filters per layer is optimised to reduce feature overlapping and complexity as compared to alternatives elsewhere in the literature. In the model, they kept pooling operations to a minimum and integrated skip connections into the network to preserve spatial information \cite{khan2021rc}.

U-Net, distinguished by its unique U-shaped architecture, has become particularly popular for medical image segmentation tasks, especially when the dataset is limited \cite{ronneberger2015u}. It leverages skip-layer connections combined with encoder-decoder multi-scale features, enhancing segmentation precision. An example of a Unet architecture is illustrated in Figure \ref{fig:unet}.    

Given U-Net's success in medical image segmentation, Livne et al. streamlined it by halving the channels in each layer, coining it ``half U-Net" for brain vessel segmentation \cite{livne2019u}. Even though there has been extensive exploration of 2D methodologies in the literature (see Tables \ref{tab:brain} especially in coronary vessel segmentation, \ref{tab:kidney}, \ref{tab:coronary}, and \ref{tab:lung}), it is observed that relying solely on 2D segmentation networks may overlook inter-slice relationships, leading to subpar segmentation \cite{wu2023multi}. Additionally, given the nonplanar nature of blood vessels, attempting to segment them without adequate volumetric information, such as on a single slice or in 2.5D scenarios, poses challenges. Although direct comparison of different studies is hampered due to the use of different datasets, the results from the studies detailed in Tables \ref{tab:brain}, \ref{tab:kidney}, \ref{tab:coronary}, and \ref{tab:lung} suggest that 3D models outperform 2D models. Specifically, 3D models achieve an average Dice score of 0.83, compared to the 0.79 average for 2D models, underscoring the significance of interslice relationships.

Consequentially, researchers have proposed 3D segmentation networks like 3D U-Net, V-Net, and VoxResNet \cite{cciccek20163d,milletari2016v,chen2018voxresnet}. Huang et al. utilised 3D U-Net with data augmentation for liver vessel segmentation and introduced a weighted Dice loss function to address the voxel imbalance between vessels and other tissues \cite{huang2018robust}. 

To enhance the segmentation abilities of the U-Net network, several modifications have been proposed. One notable advancement is the attention U-Net, introduced by Oktay et al., which is designed to suppress irrelevant regions in an input image while highlighting salient features useful for a specific task.\cite{oktay2018attention}. Bahdanau et al. originally conceived the attention mechanism to tackle the challenges arising from using a fixed-length encoding vector, which restricted the decoder's access to input information \cite{bahdanau2014neural}. This mechanism mimics human attention, enabling the model to concentrate on specific input areas while generating an output. The core idea behind this mechanism is to assign different weights to different parts of the input data, indicating how much “attention” each part should receive relative to others when producing a specific output.

Building upon the U-Net framework, Zhou and colleagues implemented U-Net ++ with the premise that the network would face a more straightforward learning challenge when the feature maps from both the decoder and encoder networks have semantic similarities \cite{zhou2018unet++}. With this goal in mind, they reconfigured the skip pathways to diminish the semantic disparity between the encoder's and decoder's feature maps. 

Sanches et al. melded 3D U-Net and Inception, dubbing it ``Uception" for brain vessel segmentation\cite{sanchesa2019cerebrovascular}. Dong and colleagues introduced a cascaded residual attention U-Net, termed CRAUNet, for a layered analysis of retinal vessel segmentation. The architecture capitalizes on the advantages of U-Net, coupled with cascaded atrous convolutions and residual blocks that are further enhanced by squeeze-and-excitation features \cite{dong2022craunet}. Drawing inspiration from U-Net and DropBlock, Guo et al. introduced the Structured Dropout U-net (SD-Unet) for coronary vessel segmentation. This design amalgamates the U-Net and DropBlock frameworks to omit specific semantic details, thereby preventing the network from overfitting \cite{guo2019sd}.In 2021, Pan et al. adopted a 3D Dense-U-Net model and replaced the standard DSC loss function with the focal loss function to tackle the issue of class imbalance in order to achieve fully automated segmentation of the coronary artery  \cite{pan2021coronary}. In 2022, Wu and colleagues introduced the Multi-Scale Interactive U-Net (MSI-U-Net), an enhancement of the 3D U-Net that enhances the precision of segmenting smaller vessels \cite{wu2023multi}. They offered a strategy for interacting with information at multiple scales, where features are transferred among vessels of varying sizes through shared convolution kernel parameters, strengthening the relationship between small, medium, and large vessels in lung CT scans.

nnU-Net \cite{isensee2021nnu} is one of the most widely utilised framework for vessel segmentation tasks, consistently delivering impressive results. It is considered the state-of-the-art for a wide variety of medical image segmentation tasks, with the original or variant of nnU-Net winning international biomedical challenges across 11 tasks\cite{gonzalez2023lifelong,isensee2021nnu}. It has also recently proposed as the network for the standardised evaluation of continual segmentation \cite{gonzalez2023lifelong}. Rather than introducing a novel model architecture, nnU-Net is designed to autonomously adapt and configure the entire segmentation framework, encompassing preprocessing, network architecture, training, and post-processing, to suit any new task. This deep learning framework is known for its flexibility, scalability, and ease of use in tackling various medical imaging segmentation problems. To this end, it is frequently used to benchmark new model approaches, or combined with other models as in ensembles to improve performance. 

It should be noted that as Moccia et al. \cite{moccia2018blood} and Garcia et al. \cite{garcia2022deep} recently mentioned, the success of the segmentation approaches is highly influenced not only by the algorithm but also by factors such as imaging modalities the presence/absence of noise or artifacts, and the anatomical region of interest. This makes direct comparisons among the studies in the literature challenging. On top of using different modalities from CTA to MRA only 29\% of the herein reviewed papers used publicly available data, further complicating efforts to replicate or compare their findings.

The data presented in Tables \ref{tab:brain}, \ref{tab:kidney}, \ref{tab:coronary}, and \ref{tab:lung} reveals a predominant use of U-Net or variants of CNNs in the majority of studies, accounting for 80\% of the cases.

\subsubsection{Generative adversarial networks (GANs)}

\begin{figure*}[!t]
\centering
  \includegraphics[width=1\textwidth]{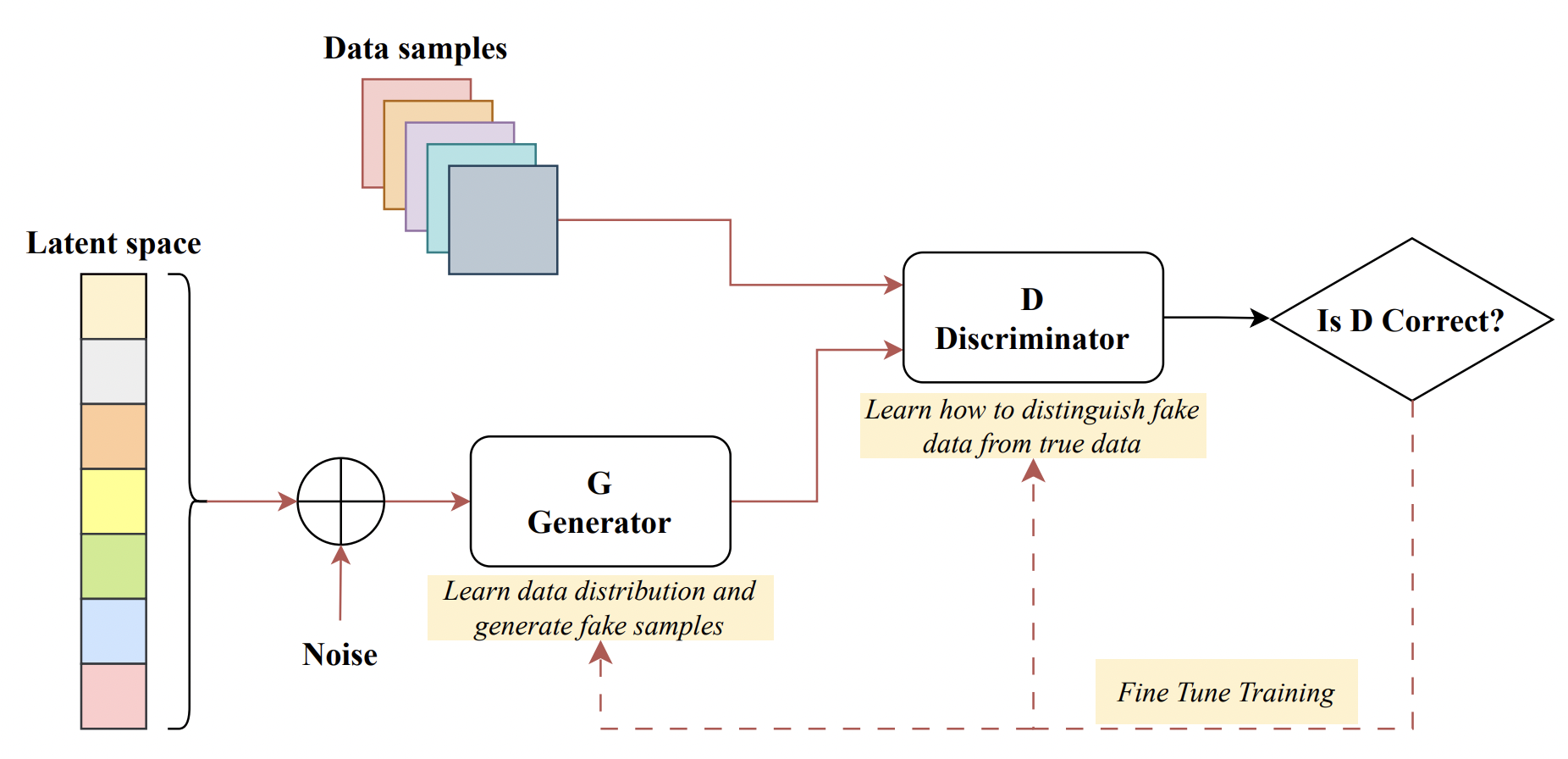}
    \caption{A sample generative adversarial network architecture - after \cite{kate2022fingan}}
  \label{fig:gan}
\end{figure*}

Goodfellow et al. introduced the Generative Adversarial Network (GAN) for synthesizing images from random noise \cite{goodfellow2014generative}. GANs represent generative models designed to approximate real data distributions, enabling them to generate novel image samples. GAN models are commonly employed for tasks such as image-to-image translation (cross-modality synthesis), image synthesis, and data augmentation. Comprising two distinct networks, a generator and a discriminator, GANs function by pitting these networks against each other during training. A sample GAN architecture can be seen from Figure \ref{fig:gan}. The generator crafts artificial images to deceive the discriminator, while the discriminator endeavours to distinguish genuine images from fabricated ones, a process termed ``adversarial training". This methodology can be adapted to train segmentation networks, where a generator is tasked with generating segmented images, and the discriminator differentiates between the predicted segmentation maps and the authentic ones. This modification prompts the segmentation network to yield more anatomically accurate segmentation maps \cite{luc2016semantic,chen2020deep}. Conditional GANs (cGANs), on the other hand, are a modified version where the generator creates images depending on specific conditions or inputs, which can be useful in vessel segmentation.

Son et al. introduced a technique employing generative adversarial training to create retinal vessel maps in the context of vascular segmentation. Their method enhanced the segmentation efficacy by employing binary cross-entropy loss during the training of the generator \cite{son2019towards}. In 2020, K.B. Park et al. introduced a novel architecture called M-GAN, which aimed to enhance the accuracy and precision of retinal blood vessel segmentation by combining the conditional GAN with deep residual blocks \cite{park2020m}. The M-generator incorporated two deep FCNs interconnected through short-term skip and longterm residual connections, supplemented by a multi-kernel pooling block. This setup ensured scale-invariance of vessel features across the dual-stacked FCNs. They integrated a set of redesigned loss functions to optimise performance, encompassing BCE, LS, and FN losses. Recently, Amran et al. introduced an adversarial DL-based model for automatic cerebrovascular vessel segmentation \cite{amran2022bv}. Their BV-GAN model utilised attention techniques, allowing the generator to focus on voxels more likely to contain vessels. This is achieved by leveraging latent space features derived from a prior vessel segmentation map, effectively addressing the issue of imbalance settings. In related research, Subramaniam et al. introduced a 3D GAN-based approach for cerebrovascular segmentation \cite{subramaniam2022generating}. While their method used GANs for dataset augmentation (generating an extensive set of examples with self-supervision), Amran et al. aimed to enhance the U-NET-based segmentation directly through the GAN technique.
Finally, earlier in 2023, Xie et al. introduced the MLP-GAN for brain vessel segmentation \cite{xie2023mlp}. This method divided a 3D brain vessel image into three separate 2D views (sagittal, coronal, and axial) and processed each through distinct 2D conditional GANs. Every 2D generator incorporated a modified skip connection pattern integrated with the MLP-Mixer block. This design enhances the ability to grasp global details.

\subsubsection{Vision transformers}

Transformers are primarily designed for sequence-to-sequence tasks but have shown significant promise in various domains, including computer vision. Vision Transformers (ViTs) are a type of neural network architecture that use transformer mechanisms to process images. First presented by Google Research in 2020, Vision Transformers have shown competitive results with traditional CNNs on image classification tasks, even surpassing them in some benchmarks. In Transformers, attention allows the model to consider other words in the input sequence when encoding a particular word, leading to the capture of long-range dependencies.
Vision Transformers (ViTs) use the same principle on image patches, allowing the model to consider distant parts of an image when encoding a particular patch. Following their success in the realm of computer vision, researchers began to explore the potential of vision transformers for medical image segmentation tasks \cite{jain2023segmenting}.

\begin{figure*}[!t]
\centering
  \includegraphics[width=0.8\textwidth]{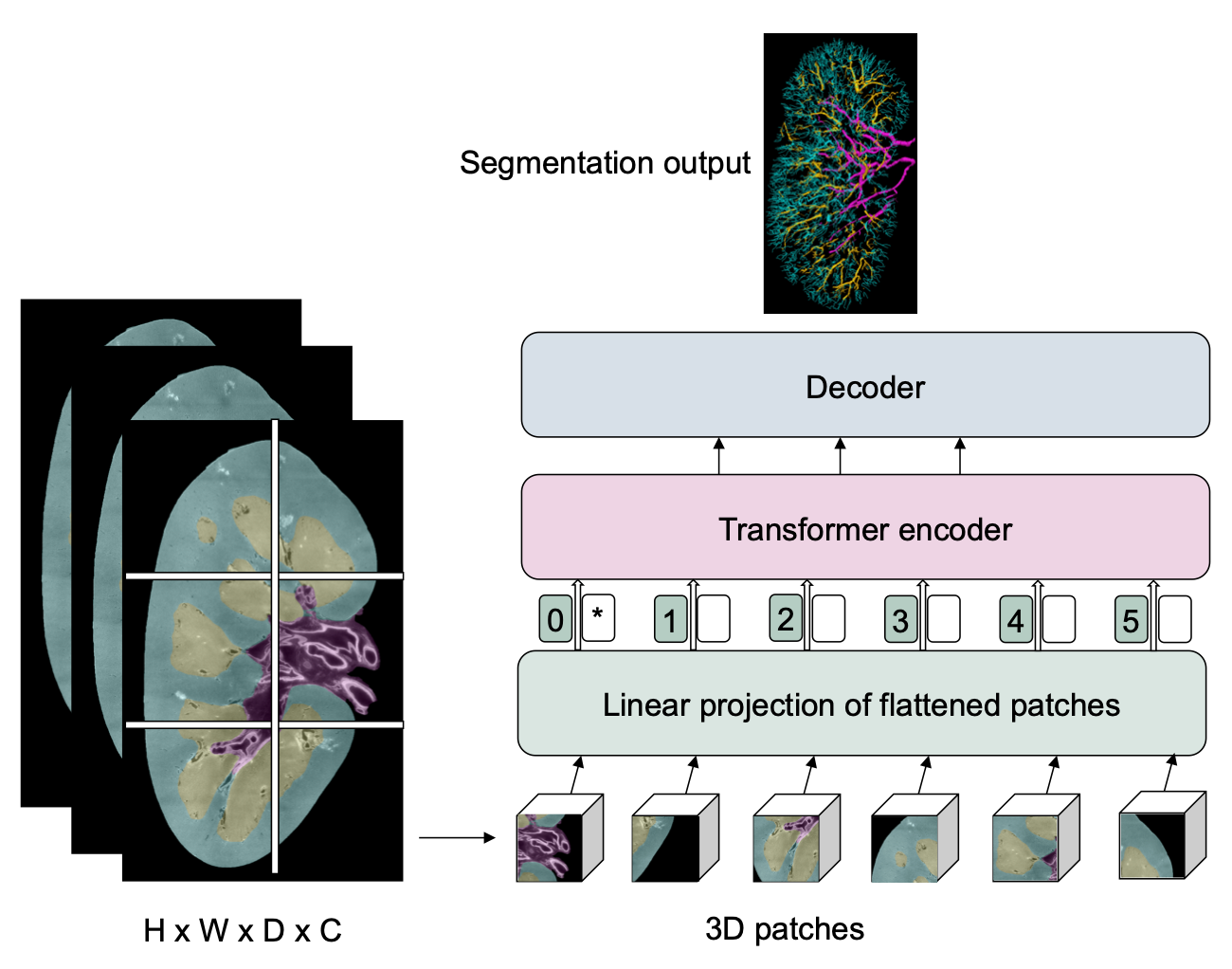}
    \caption{A sample vision transformer architecture for renal vasculature segmentation}
  \label{fig:transformer}
\end{figure*}

Instead of being processed by pixel values, images are segmented into setsized, non-overlapping sections, such as 16x16 pixels for medical image segmentation using ViTs. These sections are linearly transformed into singular vectors, a procedure called tokenization. Subsequently, to preserve the spatial context, positional embeddings are integrated with the tokenised patches. These enhanced embeddings navigate through various layers of the standard transformer encoder. To form a segmentation mask, a decoding method, which might be an upsampling layer or an alternative transformer, is applied to produce labels for each pixel in the image.  A sample vision transformer framework for vasculature segmentation is illustrated  in Figure \ref{fig:transformer}.

In 2021, Chen et al. came up with a model called TransU-Net, where a hybrid CNN-Transformer architecture is created to leverage both detailed high-resolution spatial information from CNN features and the global context encoded by Transformers by treating the image features as sequences \cite{chen2021transunet}. 

Pan et al. introduced the Cross Transformer Network (CTN), a new approach designed to understand 3D vessel features while considering their overall structure. CTN accomplishes this by combining the U-Net and transformer modules, allowing the U-Net to be more globally aware and better handle issues like disconnected or imprecise segmentation \cite{pan2022deep}.

Yu et al. introduced two new deep learning modules, called CAViT (Channel Attention Vision Transformer) and DAGC (Deep Adaptive Gamma Correction), to solve the problem of retinal vessel segmentation. CAViT combines two components: efficient channel attention (ECA) and the vision transformer (ViT). The ECA module examines how different parts of the image relate to each other, while the ViT identifies important edges and structures in the entire image. On the other hand, the DAGC module figures out the best gamma correction value for each input image. It does this by training a CNN model together with the segmentation network, ensuring that all retinal images have the same brightness and contrast settings \cite{yu2022vision}.

Zhang and his team, on the other hand,  introduced a model named TiM-Net for efficient retinal vessel segmentation \cite{zhang2022tim}. To capitalize on multiscale data, TiM-Net takes in multiscale images after maximum pooling as its inputs. Following this, they incorporated a dual-attention mechanism after the encoder to minimize the effects of noisy features. Concurrently, they utilised the MSA mechanism from the Transformer module for feature re-coding to grasp the extensive relationships within the fundus images. Finally, a weighted SideOut layer was created to complete the final segmentation.

The Swin Transformer is a modified version of the Vision Transformer (ViT), designed to further adapt the transformer structure for image-related tasks and boost its efficiency. ``Swin" gets its name from ``Shifted Window," reflecting a key feature of its design. In July 2023, Wu and his team developed the Inductive BIased Multi-Head Attention Vessel Net (IBIMHAV-Net) \cite{wu2023hepatic}. The architecture is formed by extending the Swin Transformer to 3D and merging it with a potent mix of convolution and self-attention techniques. In their approach, they used voxel-based embedding instead of patch-based, to pinpoint exact liver vessel voxels, while also utilizing multi-scale convolution tools to capture detailed spatial information.

The analysis of Tables \ref{tab:brain}, \ref{tab:kidney}, \ref{tab:coronary}, and \ref{tab:lung} indicates a less frequent application of GANs and transformers (around 20\%), suggesting that while U-Net and CNNs are the preferred methods in the field, there is still room for exploration and potential growth in the utilisation of GANs and transformers for vessel segmentation.

\section{nnU-Net for kidney vessel segmentation}\label{sec3}

From the above review of the literature, we concluded that segmentation of blood vessels from HiP-CT data should be initially attempted using nnU-Net \cite{isensee2021nnu}. As nnU-Net is one of the most widely utilised frameworks for vessel segmentation tasks and has consistently delivered impressive results, we felt that this would provide the initial baseline against which any future development of more HiP-CT-specific frameworks should be benchmarked. Therefore, we prepared a training dataset from three different human kidneys, imaged with HiP-CT and semi-manually segmented by expert annotators.  The goal was to assess how much training data is needed to provide adequate segmentation for a single kidney but also to investigate if nnU-Net trained on a subset of kidney data would generalise to a new kidney dataset given the high inter-sample variability seen between human organs. 

In this section, we will provide a  description of the HiP-CT kidney datasets, including its acquisition protocol and the segmentation process (subsection \ref{dataset}). Following that, we will present detailed information about the employed nnU-Net framework configuration (subsection \ref{architecture}), and finally, we will present the results for nnU-Net application to kidney vascular segmentation of HiP-CT data subsection \ref{evaluation}.

\subsection{Hierarchical phase-contrast tomography (HiP-CT) kidney dataset}
\label{dataset}

\begin{table}[]
\centering
\caption{The size of the kidney volumes from the dataset used in the experiments together with their gender and resolution information}
\label{tab:my-datasetinfo}
\resizebox{\textwidth}{!}{%
\begin{tabular}{cccccccc}
\toprule
\textbf{\begin{tabular}[c]{@{}c@{}}Donor \\ identifier\end{tabular}} & \textbf{\begin{tabular}[c]{@{}c@{}}Experiment \\ name\end{tabular}} & \textbf{\begin{tabular}[c]{@{}c@{}}Dataset size \\ (x,y,z) (pixels)\end{tabular}} & \textbf{Gender} & \textbf{Age} & \textbf{\begin{tabular}[c]{@{}c@{}}Scanning \\ voxel size\\ (um)\end{tabular}} & \textbf{\begin{tabular}[c]{@{}c@{}}Scan energy \\ (keV)\end{tabular}} & \textbf{\begin{tabular}[c]{@{}c@{}}Binned voxel size \\ (um)\end{tabular}} \\ \midrule
LADAF-2021-17                                                        & Kidney 1                                                            & 1303,912,2279                                                                     & M               & 63           & 25.0                                                                           & 81                                                                    & 50.0                                                                       \\
S20-28                                                               & Kidney 2                                                            & 1041,1511,2217                                                                    & M               & 84           & 25.0                                                                           & 88                                                                    & 50.0                                                                       \\
LADAF-2020-27                                                        & Kidney 3                                                            & 1706x1510,500                                                                     & F               & 94           & 25.08                                                                          & 93                                                                    & 50.16                                                                      \\ \bottomrule
\end{tabular}%
}
\end{table}

Three human kidneys were used to create the training dataset - termed kidney 1, kidney 2 and kidney 3. Kidney 1 and kidney 3 , were collected from donors who consented to body donation to the Laboratoire d’Anatomie des Alpes Françaises before death. Kidney 2 was obtained after clinical autopsy at the Hannover Institute of Pathology at Medizinische Hochschule, Hannover (Ethics vote no. 9621  BO K 2021). The transport and imaging protocols were approved by the Health Research Authority and Integrated Research Application System (HRA and IRAS) (200429) and French Health Ministry. Post-mortem study was conducted according to Quality Appraisal for Cadaveric Studies scale recommendations \cite{wilke2015appraising}.
Sample preparation and scanning protocols are described in the references \cite{brunet2023preparation, walsh2021imaging,rahmani2023micro}. Basic scan parameters and demographic information are provided in Table \ref{tab:my-datasetinfo}

\begin{figure*}[!t]
\centering
  \includegraphics[width=0.8\textwidth]{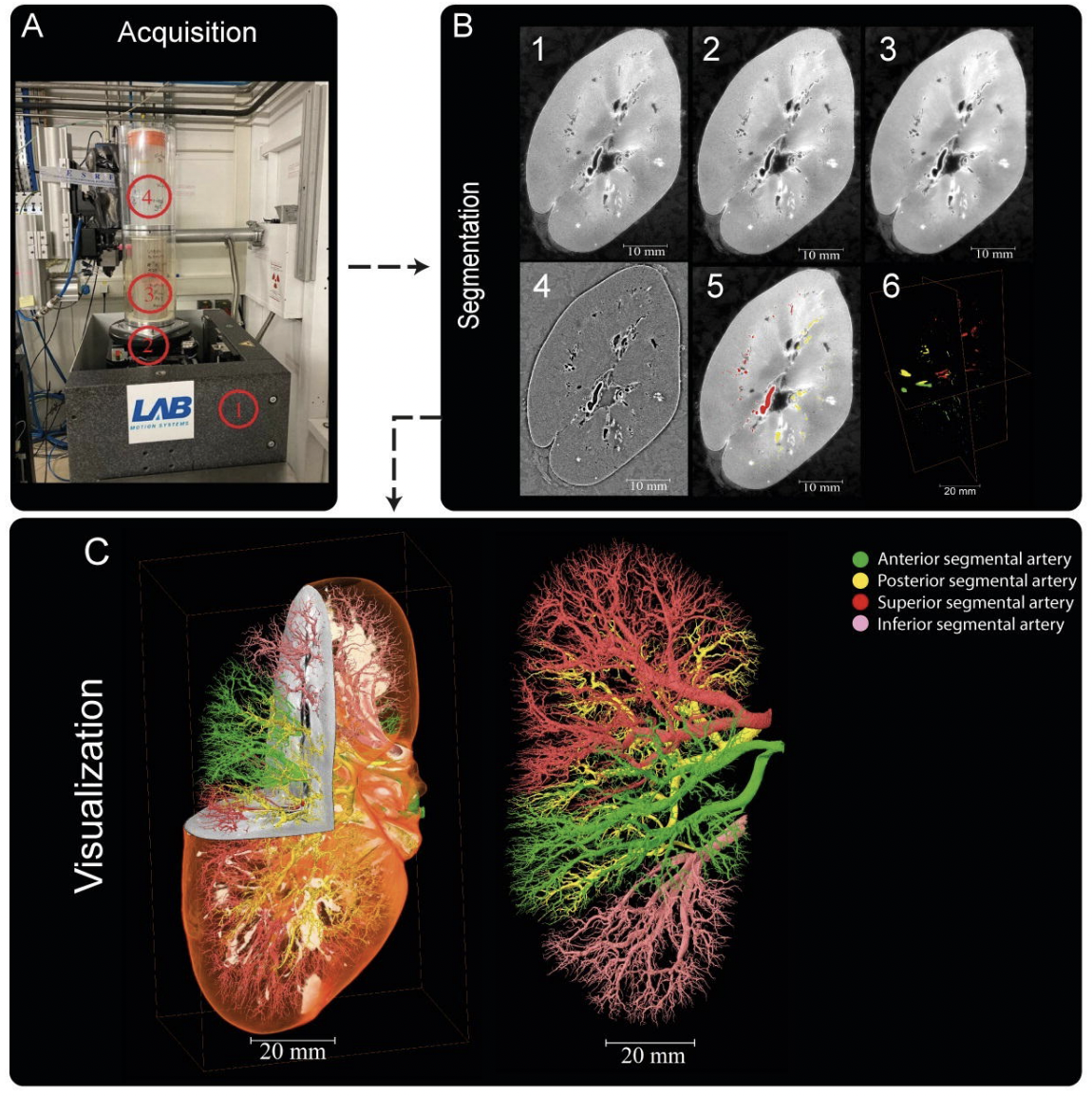}
    \caption{Processing pipeline for HiP-CT imaging and segmentation of
human kidney vasculature. (A) Setup for imaging acquisition using HiP-CT at BM05; 1) tomographic stage, 2) platform,
3) sample, 4) reference sample (B) Image processing pipeline; 1) a 2D reconstructed
image at 25 µm3
/voxel resolution; 2) binning the image by 2, 3) applying 3D median filter to
increase signal-to-noise ratio, 4) Image normalization using background detection correction,
5) Segmentation and thresholding, 6) Labelling the four main arterial branches (C) 3D
rendering of the segmented vascular network of a human kidney. Each of the main four
branching of the renal artery entering the kidney are colour-coded, Figure after \cite{rahmani2023micro}}
  \label{fig:process_pipeline}
\end{figure*}

The segmentation process of the three kidneys was carried out in Amira Version 2021.1. The reconstructed raw image data Figure\ref{fig:process_pipeline} B1 first underwent average binning x2 Figure\ref{fig:process_pipeline} B2 from the acquired resolution (~ca. 25$\mu{m}$) to ca. 50$\mu{m}$, 3D median filtering was applied in Amira-Avizo v2021.1 (3 iterations 26 voxel neighbourhood)  Figure\ref{fig:process_pipeline} B3 and filtering to visually enhance vessels appearance performed using background detection correction (Amira v2021.1; default parameter settings)Figure\ref{fig:process_pipeline}B4. Segmentation was performed in a semi-manual fashion using the Amira v2021.1 magic wand tool. This is an interactive 3D region growing tool. Using this tool, annotators select a seed voxel within a vessel in a slice (slices can be in any one of three orthogonal directions); in addition, the annotator selects and refines a combination of intensity threshold, contrast threshold and hard-drawn limits, which are used to specify the stopping criteria of the 3D region growing. In some cases, it is necessary to manually draw using a paintbrush tool in slice-by-slice locations where vessels are infilled with blood or have largely collapsed.  To provide assurance for labelling quality, an expert annotation validation process was implemented. Firstly, an independent, experienced annotator meticulously conducted a 3D proofreading of the binary labels filling in any missing vessel labels in the three orthogonal planes. After this, five randomly selected 2D circular regions of the image are selected, a third annotator marks all the labeled regions (one mark per region) (these label regions are cross sections of the vessels). The third annotator then  counts the True Positive (correctly segmented vessels),  False Negative (missed vessels), (note false positives are rare and easily removed due to the connected nature of the arterial tree). This provides a metric for what proportion of vessels have been traced or missed, and by sampling random areas in the dataset, directs the annotators to regions that may be poorly segmented. This triple validation  process was iterated to achieve final segmentations which can be accessed via \cite{kaggle}. 
It should be noted that with the binning, median filter and manual approach, vessels with diameter as small as 1-2 pixels could be segmented. Given the resolution of the images and their location within the generation vascular tree, these vessels represent interlobular arteries within the renal vascular tree and have been estimated from higher resolution HiP-CT of the same kidney, to be two branching generations from the kidney glomeruli (where the capillary bed is found). It is also of note that these samples come from a diverse age grouping 63-94yr old and have two male and one female represented. Finally it is also worthy of note that in the case of kidney 2 (S20-28) the donor's cause of death was COVID-19 which can cause microthrombi in the renal vasculature and is seen as hyperintensities particualrly around the edge of the kindey cortext, this can be seen in Supplementary movies (1-3) which show slice by slice views of each kidney.

The number of slices and size of each slice segmentation were as followed (Z, X, Y): kidney 1: (2279, 1303, 912); kidney 2: (2217, 1041, 1511); and kidney 3: (501, 1706, 1510). The network topologies automatically generated for 3D full resolution configuration for experiment 1 were a patch size (Z, X, Y)of [112,112,192], a batch size of 2  and the number of pool per axis (Z, X, Y) were [4,4,5].


\subsection{nnU-Net framework}
\label{architecture}

\textbf{Preprocessing}. Image preprocessing is a necessary step in machine learning, especially for image segmentation tasks. Notably, this step is essential in deep learning since it prepares the input data for effective model training and ensures that the model can learn meaningful patterns and relationships from the data. Proper preprocessing can improve model performance, generalisation, and robustness in handling real-world data. nnU-Net includes some automated preprocessing steps as part of its data preparation pipeline. These preprocessing steps include data augmentation (rotations, scaling, gaussian noise, gaussian blur, brightness, contrast, simulation of low resolution, gamma correction and mirroring), intensity normalization (global dataset percentile clipping, z-score with global foreground mean, z-score with per image mean), image and annotation resampling strategies (in-plane with third-order spline, out-of-plane with the nearest neighbour, third-order spline), and image target spacing (lowest resolution axis tenth percentile, axes median, median spacing for each axis). 

\textbf{Architecture}. It is important to mention that the nnU-Net framework does not have a new deep learning architecture. However, it covers the U-Net family, which has encoder-decoder architectures, including 2D U-Net \cite{ronneberger2015u}, 3D U-Net \cite{milletari2016v}, and Cascaded 3D U-Net \cite{roth2018application}. It also provides an ensemble option, which explores 2D U-Net, 3D U-Net, or 3D cascade results and chooses the best model (or combination of two) according to cross-validation performance. 

\textbf{Postprocessing}. nnU-Net framework also provides optional configuration for postprocessing on the full set of training data and annotations, including treating all foreground classes as one individual class (depending on the largest component suppression increases in cross-validation performance)

\textbf{Configuration and Training Process}. We trained the nnU-Net framework using the 3D U-Net version with 5-fold cross-validation implemented on the training sets. The evaluation results presented in the following section are the averages on testing sets of 5-fold cross-validation. We used the nn-UNet default auto-generated hyper-parameters for training, which include the learning rate \cite{chen2017deeplab} initialised as 0.01 with a polynominal decay policy of $(1 - epoch/epoch_{max})^{0.9}$, the loss function as the sum of cross-entropy and Dice loss \cite{drozdzal2016importance}, an optimizer based on stochastic gradient descent (SGD) with Nesterov momentum value of 0.99 and epoch number of 1000 with mini-batches of 250. Notably, the default nnU-Net typically requires $>30$ million parameters for the training. The proposed model was implemented in Python language\footnote{\url{https://www.python.org}} using Pytorch~\cite{NEURIPS2019_9015}. All experiments were on a NVIDIA TITAN RTX 24GB GPU.

\begin{figure*}[!ht]
\centering
  \includegraphics[width=1\textwidth]{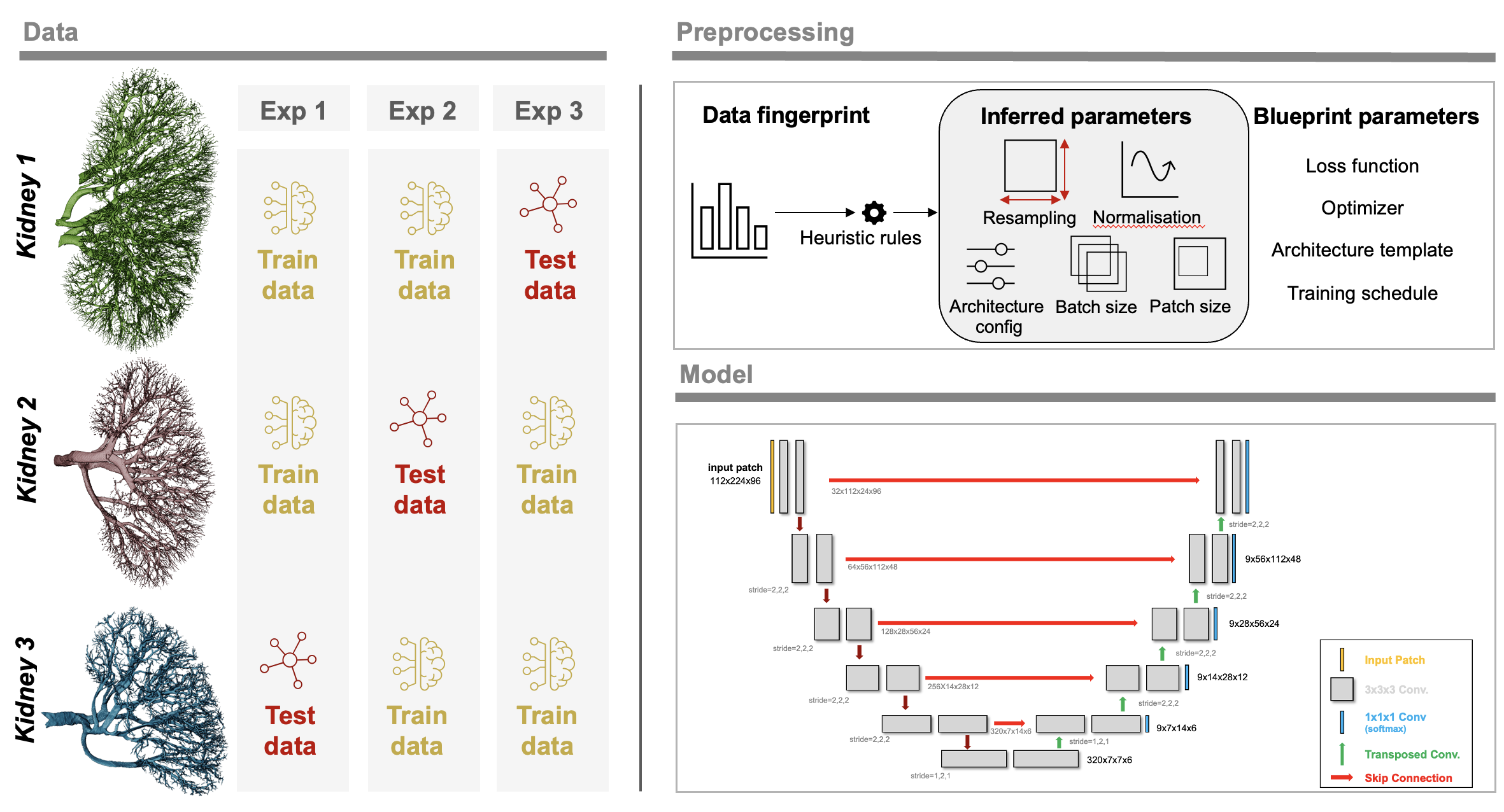}
    \caption{Overview of the experiments together with a sample 3D nnU-Net architecture }
  \label{fig:overview}
\end{figure*}

\subsection{Evaluation}
\label{evaluation}

We chose to perform four experiments. In three experiments, we used two of the kidneys as training data and one as test data. These aimed to investigate how well the nnU-Net approach would generalise to a new dataset where the donor and thus anatomy and sample prep may differ slightly; it also had the virtue of testing how different-sized training datasets impacted nnU-Net training output. The overview of the experiments can be seen in Figure \ref{fig:overview}. 

The fourth experiment aimed to see how well nnU-Net could segment the remainder of a dataset given a limited amount of labelled slices from the same dataset. Due to the size of HiP-CT datasets, such a model still has utility as segmenting vascular networks for entire organs is highly time-consuming \cite{rahmani2023micro}. In this section, we selected kidney 1 for our experimental evaluation. Half of the kidney has been utilized as a training set (1138 slices), and the remaining half as a testing set (1140 slices).

The importance of evaluation metrics in automatic image processing with machine learning cannot be emphasised enough, as they serve as a basis for determining the choice or practical applicability of a method. However, much of the research has been on creating novel image processing algorithms leaving the critical issue of reliably and objectively evaluating the performance of these algorithms largely unexplored \cite{maier2022metrics}. Moreover, some of the commonly used evaluation metrics do not always correlate with clinical applicability \cite{alidoost2023model, vaassen2020evaluation}, and the specific features of a biomedical problem may make certain metrics unsuitable, such as when the Dice Similarity Coefficient (DSC) is utilised to evaluate extremely small structures \cite{antonelli2022medical}. As a result, carefully choosing the right evaluation metric for a given problem becomes important for validating and comparing the performance of image processing methods.

In vessel segmentation, class imbalance presents a significant challenge where the pixel count for the foreground class (vasculature) is notably lesser than that of the background class (non-vasculature).

Class imbalance can hinder effective network training, as most data points are typically negative samples (or backgrounds) that often do not offer valuable learning insights. Moreover, these negative samples can dominate the positive ones (such as arteries) during the training process as loss values are predominantly generated from the negative samples.

\begin{table}[]
\centering
\caption{Results of all four experiments done using nn-UNet. SD: Surface Distance; NSD: Normalised Surface Dice; CLD: Centerline Dice; t: tolerance; Sym: symmetric}
\label{tab:my-results}
\resizebox{\textwidth}{!}{%
\begin{tabular}{@{}cccccccc@{}}
\toprule
\textbf{Expt.} & \textbf{Train data} & \textbf{Test data}     & \textbf{Dice} & \textbf{CLD} & \textbf{NSD (t=1)} & \textbf{NSD (t=0)} & \textbf{Avg. sym. SD} \\ \midrule
\textbf{1}     & Kidney 1,2          & Kidney 3               & 0.9410        & 0.8886       & 0.9651             & 0.7631             & 4.300                               \\
\textbf{2}     & Kidney 1,3          & Kidney 2               & 0.9523        & 0.8533       & 0.9518             & 0.7120             & 0.8639                              \\
\textbf{3}     & Kidney 2,3          & Kidney 1               & 0.8585        & 0.8228       & 0.8968             & 0.7132             & 2.9270                              \\
\textbf{4}     & Half of kidney 1    & Other half of kidney 1 & 0.9513        & 0.8631       & 0.9404             & 0.8549             & 2.1561                              \\ \bottomrule
\end{tabular}%
}
\end{table}

\begin{figure*}[ht]
\centering
  \includegraphics[width=1\textwidth]{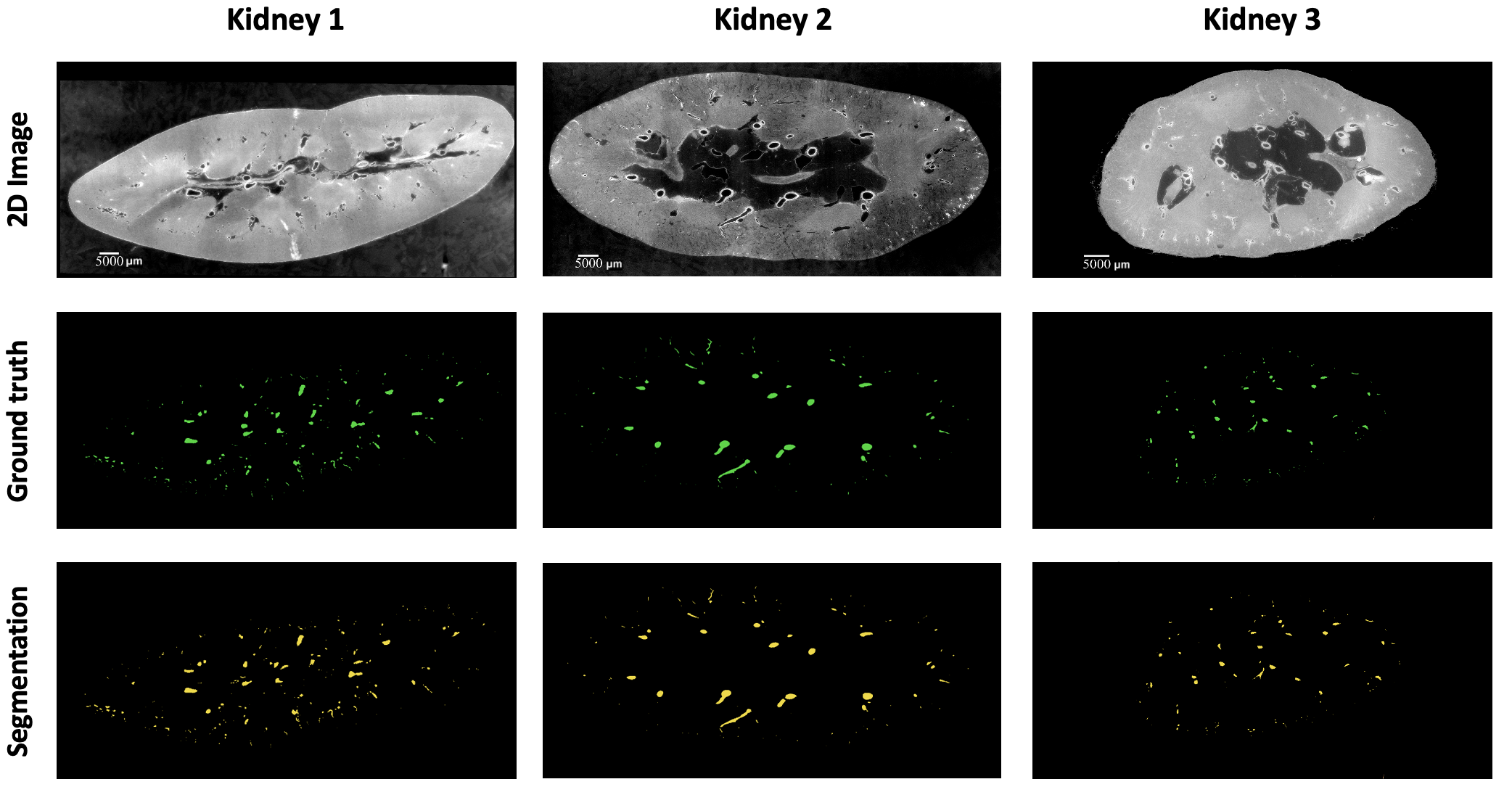}
    \caption{A sample 2D slice  from each kidney together with their corresponding ground truth labels and the output of the model segmentations (for kideny 1 the segmentation is provided by Expt. 3}
  \label{fig:2d_results}
\end{figure*}

A combination of overlap-based and boundary-based metrics is selected to evaluate different properties of the model predictions \cite{maier2022metrics}. For overlap-based metrics, Dice Similarity Coefficient (DSC or Dice) is computed and is complemented by the Centerline Dice (clDice) \cite{shit2021cldice} due to the tubular nature of blood vessels and the importance of connectivity for vascular structures. Overlap-based metrics have certain limitations such as shape unawareness and inaccurate assessment when dealing with small structures hence, boundary-based metrics, specifically Normalised Surface Dice (NSD) and Average Symmetric Surface Distance (ASSD), are also computed using MONAI v1.2.0 \cite{cardoso2022monai}. The results of the four experiments are provided in Table \ref{tab:my-results} and representative images for each kidney shown in Figure \ref{fig:2d_results}.    



\section{Discussion}\label{sec4}





In this study, we evaluated model performance for the segmentation of the arterial network from HiP-CT images of whole human kidney. Five metrics: Dice, clDice, NSD (t=1), NSD (t=0), and ASSD, with results presented in Table \ref{tab:my-results} were used to evaluate the outputs. The range of metrics we employed in our experiments provides a thorough overview of model performance and allows the state-of-the-art presented to be rigorously benchmarked against future research. For each experiment, one unseen kidney is selected as test data for evaluation of the model's generalisation capability. The 5 metrics are calculated on test data after the model training is finished. Beyond quantitative evaluations, we provided visual insights through 3D representations of the ground truth, model predictions, and false negatives. 

\begin{figure*}[!t]
\centering
  \includegraphics[width=1\textwidth]{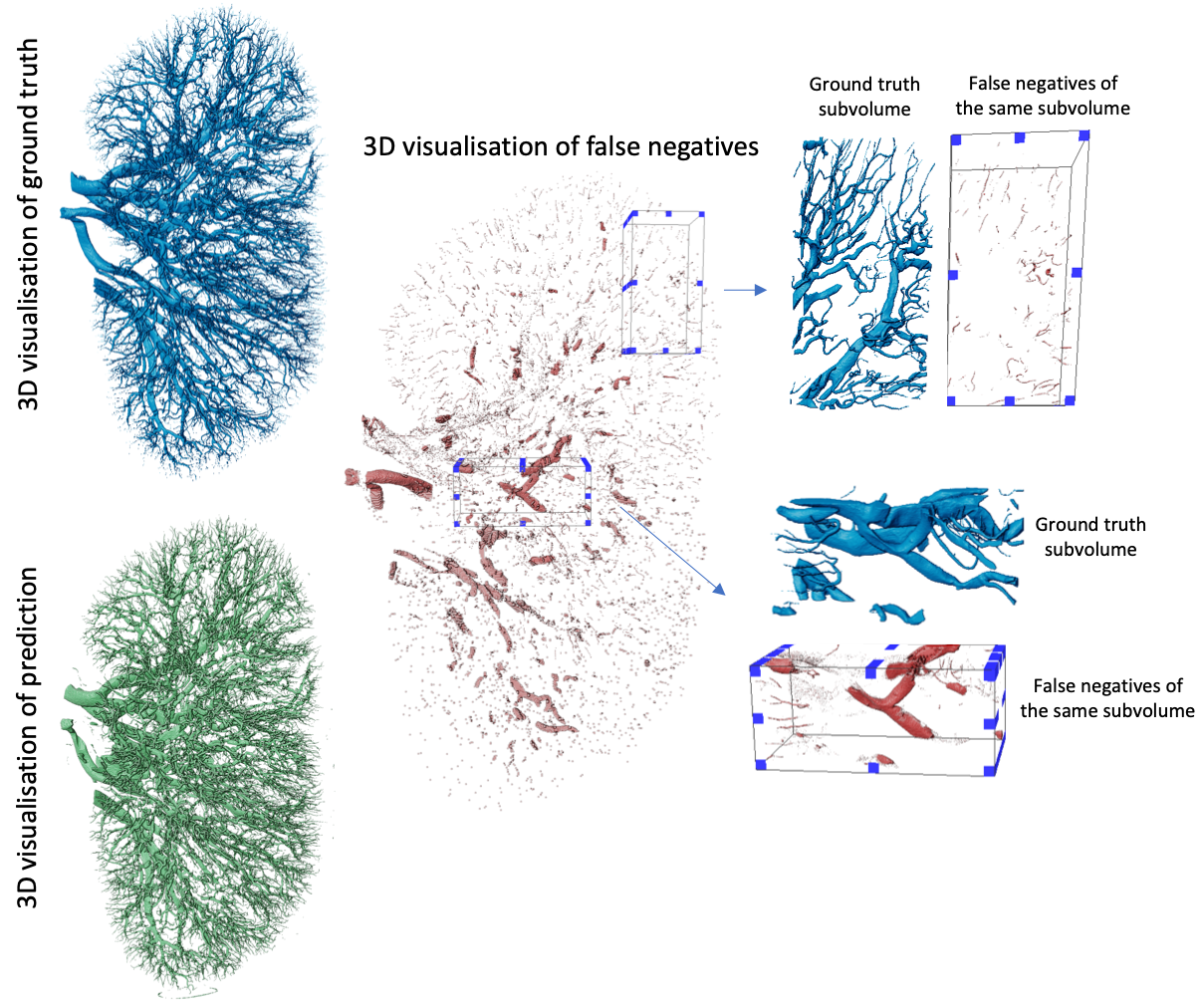}
    \caption{The visualization of  kidney 1 and model's false negative prediction together with subvolumes taken from it indicating the regions where model did perform worst} 
  \label{fig:3d_prediction}
\end{figure*}

Based on our literature review (see Tables \ref{tab:brain}, \ref{tab:kidney}, \ref{tab:coronary}, and \ref{tab:lung}), it is clear that the Dice Similarity Coefficient (DSC), also known as Dice, is the predominant metric for evaluating vascular segmentation model performance. Therefore, employing DSC for result evaluation is essential to align with existing literature, such as in \cite{hilbert2020brave}. This metric quantifies the overlap between segmentation predictions and ground truth. In our three experiments, the first and second experiments yielded superior segmentation results on the unseen kidney, with Dice scores of 0.9523 and 0.9410, respectively, whereas the third experiment attained only 0.8585. 3D examination of kidney 1, as show in Figure \ref{fig:3d_prediction}; revealed more collapsed vessels compared to kidney 1 and 3, potentially explaining the lower Dice score when the second kidney was the test subject. Nevertheless, DSC primarily assesses voxel-to-voxel concordance, overlooking several crucial characteristics of the vessels. Hence, it should not be the sole metric for deciding the performance of vascular segmentation.

Given the nature of vasculature networks - heirarchial structures of thin and elongated tubular structures, ensuring sensitivity to small vessels and mitigating boundary ambiguities is pivotal. Therefore, we also reported the distance measurement by a distance metric known as ASSD (see equation 15) to evaluate the boundary distances between predictions and ground truths. On top of ASSD, we also calculated the NSD to evaluate the similarity of the surfaces (or boundaries) of the segmented regions rather than the volumetric overlap. Instead of considering the overlap of the entire segmented regions, NSD assesses the closeness of their borders. It is particularly useful in scenarios where small boundary deviations can be critical, such as in segmenting thin structures or lesions. Tolerance ($t$) is an important hyper-parameter when it comes to NSD, which relates to the acceptable amount of deviation in the segmentation boundary in pixels. If a point on the boundary of the prediction is within this tolerance distance from a point on the boundary of the ground truth, it is considered a TP. Therefore, an appropriate NSD tolerance should usually be selected based on inter-annotator variability or some other heuristic. In our experiment, we initially set tolerance to a strict 0 pixels and relaxed it to 1 pixel afterwards, with the latter achieving around 25\% increase in the NSD score. This shows that the model fails to predict the exact boundaries of the vasculature, but 1-pixel error tolerance can highly increase the segmentation results, which indicates that while a post-processing step could be used to fix the larger vessels, the vessels that might be only 1 pixel in diameter (as is the case in our dataset) would still be missed. Consequently, models that improve segmentation of such thin vessels are needed.       

On the other hand, capturing vessel continuity and centerline of the structure is vital even if the overall vessel shape is largely captured. To this end, we applied clDice (see equation 7), a modified Dice, as another evaluation metric to our experiment. Minor deviations or inaccuracies in the centerline of vasculature structures can lead to a significant reduction of the clDice score, as it treats larger as well as thinner structures equally, thereby reducing bias toward algorithms that predict larger structures such as large arteries but miss microvasculature structures. A centerline-based metric can also provide a more consistent basis for comparison of vasculature continuity as it reduces the influence of boundary details which might be more susceptible to variations in resolution. In our case, discontinuities on the boundaries (see Figure \ref{fig:3d_prediction}) and a decrease in performance capturing smaller vessels at the end can result in lower clDice than standard Dice reported in Table \ref{tab:my-results}. Moreover, if the model underestimates or overestimates the width of the vessels the standard DSC might still be relatively high because it considers the overlap of the entire segmented region with the ground truth. However, these inaccuracies can disrupt the vessel's centerline, leading to a lower clDice score. It should be noted that since clDice depends on the underlying skeletonization algorithm, it should be carefully used, as inefficiencies or method chosen for skeletonisation can lead to inaccurate computation of the clDice score.

In Section \ref{evaluation}, we describe an additional experiment (experiment four) to evaluate the nnU-Net’s performance with a limited dataset. In this case we used just kidney 1. We trained the model on half the data and tested it on the remaining half. This experiment's training-test division was designed to test the model’s capabilities, to extrapolate from a small subset of images within a single organ to the entire dataset. This reduces the challenge of the segmentation task but also reduces the utility of the final model, however given the size of HiP-CT datasets, such an approach is likely to still have utility in particularly challenging or anatomically unique datasets. The performance metrics for the test set were Dice = 0.9513, clDice = 0.8631, NSD (t=1) = 0.9404, NSD (t=0) = 0.8549, and ASSD = 2.1561. These results indicate that nnU-Net is effective at segmenting 'unseen' kidney tissues from the same organ, reflecting its capability to learn kidney anatomy and thereby enhancing the efficiency of the labeling process. It also shows the expected improvement by comparison to the other three experiments, indicating the challenges with accounting for anatomical variation across human samples, or variations in the data aquisition and data reconstruction parameters see \cite{brunet2023preparation,walsh2021imaging,xian2022multiscale} for details on protocol variations. The variation is also expected as the HiP-CT technique continues to develop and as the Human Organ Atlas extends it would be possible to use automated segmentation techniques to perform large studies to analyse vascular network differences due to pathology, age or sex etc of the donor.


From the discussion on metrics above, the benchmarking methodology of nnUNet fails in several scenarios when applied to HiP-CT data on vascular segmentation. With whole organ HiP-CT we are able to manually segment and create training data down to the level of the interlobular arteries or approx. two generations from the capillary bed. These arteries have a minimum radius of ca. 40$\mu{m}$ \cite{rahmani2023micro}, (1-2 pixels in diameter). However, the experiment scores reveal that the failed prediction on such small vessels leads to discontinuity problems for the vasculature structure estimation. The work in \cite{shit2021cldice} has shown an improved continuity performance on vasculature segmentation with clDice as part of the loss function when training the network. Integration of rich hierarchical representations of thin and long structures by use of spatial attention and channels attention modules also can potentially improve the outcomes \cite{mou2021cs2}. Another challenge that may hinder vasculature continuity is edge identification and segmentation. Considering the imbalance of edge and non-edge voxels, an edge-reinforced neural network (ER-Net)  \cite{xia20223d} combining a reverse edge attention module \cite{zhang2020cerebrovascular}  and an edge-enforced optimisation loss to discover spatial information of the edge structures could potentially increase the vasculature continuity on segmentation results. In both the small vessel and edge enhancement cases it is important to remember the HiP-CT is a propagation-based phase contrast technique and thus has intrinsic contrast for small structures and edges between structures. Tomographic reconstruction is a part of the process of HiP-CT data volume data production, and can have a large impact on the ease of segmentation. Tuning the tomographic reconstruction pipeline to enhance specific features, whilst challenging could be a powerful approach to segmenting specific smaller structures. 

Relative to the ground truth, the benchmark model appears to overlook certain vasculature, particularly struggling to identify certain large vessels Figure \ref{fig:3d_prediction}. Upon closer examination of the pertinent sub-volumes, it becomes evident that these misdetected large vessels correspond to collapsed vessels within the organs. As shown in Figure \ref{fig:3d_prediction} of the prediction on kidney 1,  collapsed vessels have a  flat appearance, morphologically different from the noncollapsed vessels seen elsewhere.  This suggests that the model’s performance falters notably in the presence of disrupted structures, and these could be the potential cause for the observed lower scores when kidney 1 is used as the test set.
      
Therefore, HiP-CT scans in high-resolution present challenges to deep learning methods applied to different tasks such as segmentation. These scans highlight more details that must be captured by the models. Using a larger patch size for training can ease the problem. However, there is a trade-off between the GPU memory and the size of the receptive field. As a result, integrating regional features and their global dependencies remains a research direction.     


\section{Conclusion and future work}\label{sec6}

The vascular system is vital for nurturing and supporting every organ in the body, and its abnormalities can be indicative of various diseases. Developing accurate and automated segmentation of vasculature, that can be applied to nasent imaging technologies such as HiP-CT, could pave the way for scalable vascular segmentation across large-scale datasets. Scalability in turn enables quantification across a large demographic population, therefore assisting the construction of a data-driven Vascular Common Coordinate Framework \cite{boppana2023anatomical,borner2021anatomical} for human atlasing projects such as the Human BioMolecular Atlas Program \cite{hubmap2019human}, the Human Organ Atlas Project \cite{walsh2021multiscale} and the Cellular Senescence Network Program   \cite{sennet2022nih}. 

This paper offers a comprehensive review of recent literature focusing on deep-learning techniques for blood vessel segmentation. We primarily delve into the recent deep-learning methodologies, emphasizing the challenges associated with vasculature segmentation. Our aim with this review is to lay a solid foundation for researchers, building robust models for vessel segmentation, especially using phase contrast imaging.

For these models to be widely accepted in the medical field, we need to address certain gaps in the current research. From our literature survey, we deduce the following:
(i) Publicly accessible datasets for vascular segmentation are limited. Even within private datasets, the size available for training is often limited.
(ii) Labeling the vascular region remains a formidable challenge and requires extensive effort.
(iii) Most studies employ private data and provide internal validation. This makes direct comparisons between studies challenging and prevents the assessment of algorithm efficacy.
(iv) Every phase of the imaging process; from sample preparation, imaging, modality specific artefact correction, pre-processing and algorithm optimization to post-processing, holds immense importance. The post-processing stage, especially, has a great impact on the final segmentation. 

Many studies focus on mapping out complex blood vessel networks, mostly concentrating on larger vessels yet the significance of smaller vessels, which are frequently overlooked, cannot be understated from a (patho-)physiological point of view \cite{ackermann2022bronchial,ackermann2022fatal}. Moreover, the choice of evaluation metrics and loss functions in vessel segmentation is often unclear, calling for clearer guidelines. Through our study, we aim to assist researchers in identifying the most suitable metrics for their analyses. One significant challenge with vessel segmentation models is their limited generalisability, as most studies rely on private datasets. To enhance adaptability, models should be cross-validated against diverse data. Providing both more public datasets and pre-trained model weights could benefit future research. By making our data publicly accessible, through a Kaggle competition (https://www.kaggle.com/competitions/blood-vessel-segmentation/data) and following its conclusion, through the human organ atlas portal (www.human-organ-atlas.esrf.eu), we invite researchers to evaluate their algorithms on HiP-CT data, fostering advancements in the field.

\section*{Acknowledgement}

The authors would like to express their gratitude to P. Masson (LADAF) for dissections of body donors', the Hannover Biobank for provision of donor organs,Emer O'Leary for segmentation input, E. Boller, K, Dollman, C. Berruyer and C. Jarnias for their help in setup, developments, improvements and imaging. The authors would like to express gratitude for the financial support provided by the Chan Zuckerberg Initiative DAF (2020-225394), an advised fund of SVCF, the MRC (MR/R025673/1), and ESRF beamtimes (md1252 \& md1290). P.D.L. is supported by a Royal Academy of Engineering Chair in Emerging Technologies (CiET1819/10). This research was funded in part by
the Wellcome Trust [209553/Z/17/Z].
The grant of the European Research Council (ERC); European Consolidator Grant, XHale to Danny Jonigk (ref. no.771883). This work was supported by the German Registry of COVID-19 Autopsies (DeRegCOVID, www.DeRegCOVID.ukaachen.de; supported by the Federal Ministry of Health—ZMVI1-2520COR201), and the Federal Ministry of Education and Research as part of the Network of University Medicine (DEFEAT PANDEMIcs, 01KX2021).

This research has been funded by the NIH Common Fund through the Office of Strategic Coordination/Office of the NIH Director under awards OT2OD033756 and OT2OD026671, by the Cellular Senescence Network (SenNet) Consortium through the Consortium Organization and Data Coordinating Center (CODCC) under award number U24CA268108, by the Kidney Precision Medicine Project grant U2CDK114886, by the NIDDK under awards U24DK135157 and U01DK133090 and by The Multiscale Human CIFAR project. The funders had no role in study design, data collection and analysis, decision to publish, or preparation of the manuscript.

\bibliography{sn-bibliography}


\end{document}